\documentclass[remotesensing,article,accept,moreauthors,pdftex,10pt,a4paper]{Definitions/mdpi}
\firstpage{1}
\pubvolume{1}
\issuenum{1}
\articlenumber{0}
\pubyear{2021}
\copyrightyear{2021}
\externaleditor{\hl{Academic Editor:} %MDPI: please add Academic Editor.
  }
\datereceived{}
\dateaccepted{}
\datepublished{}
\hreflink{https://dx.doi.org/} % If needed use \linebreak
\usepackage{makecell}
\usepackage{pdflscape}
\usepackage{float}
\Title{\highlighting{Review} %MDPI: the article type in the document is different with that in Susy.
 on the Role of GNSS Meteorology in Monitoring Water Vapor for Atmospheric Physics}

\TitleCitation{Review on the Role of GNSS Meteorology in Monitoring Water Vapor for Atmospheric Physics}

\Author{\hl{Javier Vaquero-Martinez} %Please carefully check the accuracy of names and affiliations.  these author name are different from redmine, please note
 $^{1,2,}$*\orcidA{} and \hl{Manuel Anton} 
 $^{1,2}$\orcidB{}}

\AuthorNames{Javier Vaquero-Martinez and Manuel Anton}

\AuthorCitation{Vaquero-Martinez, J.; Anton, M.}

\address{%
$^{1}$ \quad Department of Physics, \hl{Universidad de Extremadura;} %MDPI: please add city, post code and country.
  mananton@unex.es\\
$^{2}$ \quad Instituto Universitario de Investigacion del Agua, Cambio Climatico y Sostenibilidad (IACYS), \hl{\mbox{Universidad de Extremadura}} %MDPI: please add city, post code and country.
}
\corres{Correspondence: javier\_vm@unex.es}

\abstract{After 30 years since the beginning of the Global Positioning System (GPS), or more generally Global Navigation Satellite System (GNSS) meteorology, this technique has proven to be a reliable method for retrieving atmospheric water vapor; it is low-cost, weather independent, with high temporal resolution and is highly accurate and precise. GNSS ground-based networks are becoming denser, and the first stations installed now have quite long time-series that allows for the study of the temporal features of water vapor and its relevant role inside the climate system. In this review, the different GNSS methodologies to retrieve atmospheric water vapor content are examined, such as tomography, conversion of GNSS tropospheric delay to water vapor estimates, analyses of errors, and combinations of GNSS with other sources to enhance water vapor information. Moreover, the use of these data in different kinds of studies is discussed. For instance, the GNSS technique is commonly a reference tool for validating other water vapor products (e.g., radiosounding, radiometers onboard satellite platforms, ground-based instruments). Additionally, GNSS retrievals are largely used in order to determine the high spatio-temporal variability and long-term trends of atmospheric water vapor or in models with the goal of determining its notable influence on the climate system (e.g.,~assimilation in numerical prediction, as input to radiative transfer models, study of circulation patterns, etc.).}

\keyword{GNSS; GPS; water vapor; ground-based; methodology; validation; spatio-temporal; time-series; meteorology; climate}

\begin{document}

%%%%%%%%%%%%%%%%%%%%%%%%%%%%%%%%%%%%%%%%%%
\hypertarget{introduction}{%
\section{Introduction}\label{introduction}}

Water vapor is a trace gas of the Earth's atmosphere. Despite its relatively low
concentration, it is of paramount importance to the Earth's climate. Firstly,
water vapor is the major contributor to the natural greenhouse effect and,
therefore, in global warming, presents positive climate feedback
\citep{Colman2003, Colman2015}. It also has an obvious role in the hydrological
cycle, and plays a key role in energy transport. Water is evaporated at low
latitudes, and its vapor is transported to higher latitudes where, by
condensation, it releases high amounts of heat \citep{Myhre2013}.

Water vapor can be quantified using different methods. One of the most common is the
integrated water vapor (IWV), which is also known by other names in the literature,
such as precipitable water vapor (PWV), precipitable water (PW), integrated
precipitable water vapor (IPWV), and so forth. This quantity represents the integral of
the concentration of water vapor along the vertical path, and can also be
pictured as the height that, in a vessel of a unit cross-section, the water vapor
in a column of the same cross-section would reach if all of it condensed to
liquid water. The density of liquid water allows the establishment of the equivalence
between the superficial concentration (in \(\mathrm{kg~m^{-2}}\)) and the height
in the vessel (in \(\mathrm{mm}\)).

Another interesting measurement of water vapor is its distribution with height.
This distribution can be given as a mixing ratio (grams of water per gram of dry
air), as relative humidity, or other measurements of concentration.

Global Navigation Satellite Systems (GNSS) allow the retrieval of information about
atmospheric water vapor, with different techniques. In this article, we will
focus on the ground-based techniques, since GNSS radio-occultation techniques
have already been discussed elsewhere \citep{Bonafoni2019, Liu2017, Wickert2009}.

Global Positioning System (GPS) meteorology started when
\citet{Bevis1992a} suggested this approach to remotely sensing water vapor, in a time
when even GPS, the first GNSS constellation, was not still completed. In that
paper, the methodology to derive IWV from tropospheric delay is thoroughly explained. Moreover,
some possible applications are depicted. One of them is the obvious use of
GNSS-receiver networks to monitor IWV, but also the idea of water vapor
tomography by using dense networks is shown.

The main idea behind remote sensing of the water vapor through GNSS is that the
GNSS signal between satellite and ground-based receiver is delayed by the
troposphere, along with other effects. The GNSS processing for obtaining the
position of the receiver needs to account for this, which is called
slant tropospheric delay (STD). This is usually given in terms of the zenith
tropospheric delay (ZTD), since it is not satellite-dependent and, to convert
STD to ZTD, a mapping function is needed. Once ZTD is obtained, meteorological
data (typically surface atmospheric pressure) are used to compute the non-dipolar
component of this ZTD, known as the zenith hydrostatic delay (ZHD). By
subtracting the ZHD from the ZTD, only the dipolar component remains, which is
exclusively due to the presence of water vapor (because it is the only
atmospheric component with permanent dipolar moment). This component is called
the zenith wet delay (ZWD). It can be converted to IWV using a factor usually
represented by \(\Pi\) (sometimes by Q). The \(\Pi\) factor has been calculated by
many methods in the literature (see Section
\ref{conversion-of-tropospheric-delays-to-water-vapor}). The most accurate methods model \(\Pi\) as a function of the weighted mean temperature of the atmosphere, also known as Davis temperature (\(T_m\)).

One simple way to estimate \(T_m\) is through linear regression with surface
temperature (\(T_s\)). \citet{Bevis1994} performed a linear regression of \(T_m\) and
surface temperature (\(T_s\)) using radiosonde data in the United States, which is
currently known as the Bevis Formula (\(T_m =70.2+0.72\cdot T_s\)). This work also
included a more detailed explanation of the method for ground-based GNSS water
vapor monitoring, providing specific figures for all the constants involved, and
discussing their uncertainty in depth. The refractivity constants were obtained
by statistical treatment of values from other studies, while the computation of
\(T_m\) values was also carried out using numerical weather prediction models
(NWP), yielding a root mean square error (RMSE) of \(2.4~\mathrm{K}\) (as compared
to the RMSE of \(4.7~\mathrm{K}\) from using the so called ``Bevis formula'').

\begin{figure}[H]

\includegraphics{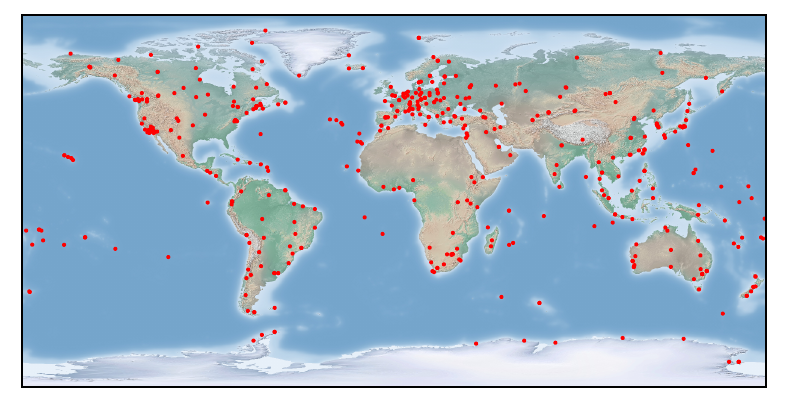}
\caption{\label{fig:igs}Current network of 506 stations belonging to IGS. Color and shape are related to the last data available date. Source: \url{https://www.igs.org/maps/\#station-map} \hl{(accessed on).} %Please add accessed date.
}
\end{figure}

After these studies, the importance of water vapor and the need of improving its
spatial and temporal resolution and coverage was acknowledged in some studies, such as 
\citep{Dabberdt1996}. Moreover, GNSS meteorology was found to be suitable for this
task, as most of the infrastructure was already there (for navigation and
geodetic applications, mainly). Therefore, it was a unique opportunity to obtain
more meteorological information at a very low cost. However, these efforts would
be useless without the information provided by services like the International
GNSS Service IGS \citep{Dow2005}, operating since 1994 and providing necessary
information for the processing of GNSS data, such as precise and real-time orbits,
clock corrections, tropospheric products, and so on. The IGS counts with a global network of
ground-based receivers and a set of Analysis Centers to process their
data. This global network is depicted in Figure~\ref{fig:igs}.

It must be noted that the processing of tropospheric delays must be finely tuned
for meteorological purposes. For navigation techniques, the estimation of the different delays do
not need to be very precise, as long as their combination adds up to the total
delay suffered by the satellite-receiver signals. However, for meteorological
purposes, the other delays need to be obtained with high precision to avoid that part
of the error can be transferred to the tropospheric delay estimate. For
instance, the ocean tide loading can induce errors of up to \(1~\mathrm{cm}\) in
ZTD if it is not properly modeled \citep{Dach2000}.

\textls[-5]{GFZ-Potsdam started the development of one of the first systematic near
real-time water vapor products based on a German network and in association with
other research centers. This was the so-called GNSS Atmosphere Sounding Project
(GASP). It was partly dedicated to space-based GNSS observations but also
to ground-based measurements} \citep{Dick2001}. The STD started to be
assimilated in analysis experiments for dense GNSS networks, developing ways of
estimating the 3D structure of water vapor \citep{MacDonald2002}. The use of
ground-based GNSS networks for tomography of the troposphere, obtaining 4D maps of
water vapor, was first discussed by \citet{Flores2000}. They proposed a methodology
(software LOTTOS) to retrieve these maps from STD measurements
and concluded that the technique could be used for
monitoring the troposphere.

Later, \citet{Morland2006} presented the STARTWAVE (STudies in Atmospheric Radiative
Transfer and Water Vapour Effects) project, whose objective was to obtain a
database of different water vapor products (including ground-based GNSS ones)
that allowed the study of the role of water vapor in the climate. At the same
time, the idea of using GNSS networks for water vapor tomography was still in the
air, and \citet{Bender2007} considered the idea of different network geometries.
They concluded that the spatial resolution would be given by the mean
inter-station distance. It was also reported that the quality and quantity of
measurements depend on the distribution of the satellite constellation and,
therefore, on time, they would improve as more satellites (including other constellations
such as GLONASS or Galileo) became available. Multi-GNSS processing became
operational for near real-time applications a few years later, when \citet{Li2015}
first compared this kind of measurement with single-GNSS ones. The processing
was possible thanks to the efforts of the IGS Multi-GNSS Experiment MGEX \citep{Montenbruck2017}.

This article reviews different methodologies used to retrieve water vapor
information from GNSS processing in Section~\ref{met}. The vast applications
of these processing techniques are discussed in the following sections, namely
the inter-comparisons with other techniques (Section~\ref{ic}), the
spatio-temporal analysis of water vapor distribution (Section~\ref{sp}), and
the impact of water vapor variability on meteorology and climate (Section~
\ref{impact}). Finally, conclusions are drawn in Section~\ref{conclusions}.

\hypertarget{met}{%
\section{GNSS Methodologies}\label{met}}

\hypertarget{conversion-of-tropospheric-delays-to-water-vapor}{%
\subsection{Conversion of Tropospheric Delays to Water Vapor}\label{conversion-of-tropospheric-delays-to-water-vapor}}

The conversion of ZWD to IWV is performed through the factor \(\Pi\). Typically,
its value is estimated by the relation described by Equation \eqref{eq:pi},
\begin{equation}
 \Pi = \frac{10^6}{\rho R_v \left[k_3/T_m + k'_2\right]},
 \label{eq:pi}
\end{equation}
where \(\rho\) is the density of liquid water, \(R_v\) the specific gas constant for
water vapor, \(k_3\) and \(k'_2\) are refractivity constants and \(T_m\) is the
weighted mean temperature of the atmosphere, defined as Equation \eqref{eq:tm}:
\begin{equation}
 T_m=\frac{\int \left(P_v/T\right) \mathrm{d}z}{\int \left(P_v/T^2\right) \mathrm{d}z}.
 \label{eq:tm}
\end{equation}

\(T_m\) is typically obtained by using a linear regression between surface
temperature (\(T_s\)) and \(T_m\) (obtained from radiosondes). The relationship \(T_m - T_s\) given \mbox{by \citet{Bevis1992a}} was obtained using radiosondes in the United States.
Logically, some efforts have been made to improve it for application to other
locations. \citet{Emardson2000} considered two models for \(\Pi\) in Europe: (1) series
expansion of the deviation from mean surface temperature, and (2) a combination
of the former with a seasonal variation based on decimal day of year and
latitude. Both models presented RMSE values between \(1.0\%\) and \(1.2\%\), fitting the
coefficients to all Europe, by regions or by site. Slight increases in RMSE
values were found as the fitting involved wider regions. In more equatorial
regions, such as southern India, \(\Pi\) has been reported to be less dependent on
\(T_m\) variations, allowing to set it to a fixed value without relevant precision loss
\citep{SureshRaju2007}. Another work addressed the model to calculate \(T_m\) in tropical regions, finding that the Bevis formula is shown to be less adequate with high temperatures and low variability. In fact, the definition of \(T_m\) assumes that temperatures are not so high, and \citet{Bevis1992a} indicates a larger dispersion for high \(T_s\) values. Therefore, it is suggested that the redefinition of \(T_m\) should be addressed for tropical regions.

NWP, as with the reanalysis from European Centre for Medium-Range Weather Forecasts
(ECMWF), have been widely used to retrieve surface pressure and \(T_m\) since 2009
approximately. \citet{Heise2009a} obtained a global 5-minute data-set from the IGS GNSS
network using this approach. The ECMWF reanalysis version used (ERA-Interim) had
6-h  resolution, and the estimated uncertainty due to time interpolation was
less than \(0.2~\mathrm{mm}\). Later, \mbox{\citet{Wang2019d} compared} the GNSS IWV product
using the new ECMWF reanalysis (ERA5) for pressure and \(T_m\), on the one hand, and using the legacy
version (ERA-Interim), on the other hand, against a reference GNSS IWV product using data from
meteorological stations. The biases and RMSE of both products were of a few
tenths of a millimeter. Against radiosondes, the RMSE was of about 2--4~ \(\mathrm{mm}\). ERA5 interpolation data-set was found to be more stable than
ERA-Interim one. \citet{Zhang2019d} showed that ERA5 GNSS IWV improved ERA-Interim GNSS
IWV in China, thanks to ERA5's pressure enhancement, and with a similar performance on
\(T_m\). The contribution of ERA5 to the IWV error was estimated to be less than
\(1~\mathrm{mm}\), and diurnal variations were better caught by ERA5 GNSS IWV than
by ERA-Interim GNSS IWV.

Moreover, there have been some approaches to use NWP data for retrieving mapping
function parameters, pressure and temperature \citep{Bohm2015, Landskron2018}. These approaches
can be useful when there are not any meteorological stations available nearby the GNSS
station or when access to NWP data is difficult.

Another possible approach is to use local models of
temperature and pressure to convert ZTD to IWV, using local meteorological
stations over a long time series. \citet{Charoenphon2020a} followed this methodology in
Thailand, finding that RMSE was below \(3 ~\mathrm{mm}\), therefore it was
considered suitable for weather nowcasting.

All of these different methodologies to convert ZTD to IWV have
advantages and disadvantages. The most precise approach is to use measurements of
temperature and pressure at the station location, but sometimes this is not
possible or not fast enough for some applications. In addition, it is important to
apply the necessary corrections when the meteorological station and the GNSS
receiver are not at the same height. Moreover, the use of models can be very
helpful for some applications or to retrieve data from receivers that are far
away from meteorological stations or to extend the time series of stations that
have a meteorological station nearby in the present but not in the past.

\hypertarget{error-analysis}{%
\subsection{Error Analysis}\label{error-analysis}}

\textls[-8]{Since GNSS meteorology is a rather indirect technique, there can be sources of error
in each step of the process. For instance, the presence of patterns of water vapor
gradients can induce errors in GNSS-derived IWV to the point that
methodologies have been developed to correct for this \citep{Iwasaki2000}. Another
important feature when computing ZTD from GNSS measurements is the satellite
Antenna Phase Center (APC) variations. The APC model changed with the addition
of new satellites to the GNSS constellation.} \mbox{\citet{Jarlemark2010} studied} the influence
of not applying proper modeling for APC, obtaining spurious, additional trends
in IWV of up \(0.15~\mathrm{mm~year^{-1}}\). On 6 November   2006, IGS changed
their processing strategy from using relative antenna phase center calibrations to
absolute ones. \mbox{\citet{OrtizdeGalisteo2010} showed} the impact of this change in GNSS IWV
products, using data in four locations in Spain from two years before and after
this change. With relative calibrations, a 2--3~\(\mathrm{mm}\) bias of GNSS with
respect to other instruments (radiosonde, sun-photometer) was found. However,
GNSS IWV after the change to absolute calibrations exhibited a practically zero
bias. In addition, RMSE reduced to half and correlations \mbox{slightly increased.}

Satellite positions and clock corrections are fundamental to the retrieval of
tropospheric GNSS products. However, precise data can only be obtained
after a few days, so they cannot be used for near real-time applications. A comparison of
near real-time water vapor product with final products to
compare the effect of rapid/final satellite orbit estimation obtained
negligible bias and RMSE below \(1~\mathrm{mm}\) \citep{Yadav2010}.

The mapping functions that transform slant delays into zenith delays are also
error-prone. \citet{Labib2019} compared GNSS IWV computed with Global Mapping Function
(GMT) and Vienna Mapping Functions 1 (VMF1) on a global scale.
Both products were compared against radiosonde observations. The authors concluded that
there was not a better option. However, both mapping functions must be
revisited for climate zones other than mid-latitude regions.

Another feature in ZTD processing that can affect IWV measurements is cutoff in
elevation angle. \citet{Ning2012} computed IWV trend using data-sets obtained with 8
different cutoff elevation angles at seven sites in Finland and Sweden for 14
years (1997--2010). In comparison with radiosonde trends, best RMSE was found for
angles of \(10^\circ\) and \(15^\circ\). However, the best correlation (\(0.88\)) was
found for cutoff angles of \(15^\circ\). Therefore, it was shown that GNSS data
contained information
about trends, but these were too small to be uniquely detected.

It is important to notice that, specially when dealing with trend estimation,
GNSS-IWV time series data must be retrieved with the same parameters to avoid
inhomogeneities in the series.

\hypertarget{tomography}{%
\subsection{Tomography}\label{tomography}}

Although there have been some experiments with single-station retrievals of the vertical structure of water vapor \citep{Caputo2000, Iassamen2009}, profiles are generally obtained from network solutions. Otherwise, a distribution has to be decided a priori (typically a decreasing exponential with height). One of the first efforts to water vapor tomography was made by \citet{Bi2006a}, who developed a method for obtaining slant water vapor (water vapor integrated along the slant path) and compared it to water vapor radiometer (WVR), with RMSE less than \(4~\mathrm{mm}\).

\begin{figure}[H]
\includegraphics[width=2.13in]{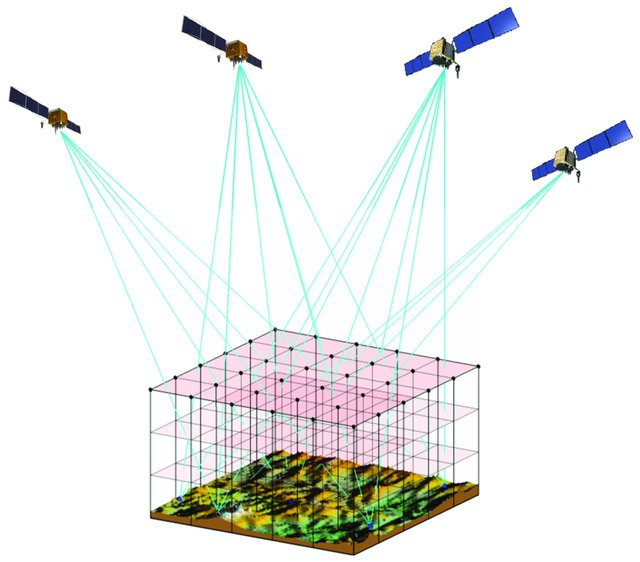} \caption{Voxel-based representation of tomography, with signal rays crossing the voxels. Source: \citet{Dong2018}.}\label{fig:voxel}
\end{figure}

Voxel approaches, however, discretize the space in cubes (voxels) and try to
obtain the concentration of water vapor in each of these voxels. This way, they
provide a 3D (or 4D, if time is also considered) representation of water vapor.
There are, however, many different approaches to accomplish this and none of
them are trivial. One way to optimize the process was provided by \citet{Yao2017}.
They proposed the use of a long-term mean radiosonde water vapor profile in the
location of the GNSS network. The horizontal voxel definition was optimized using
a non-uniform symmetrical division, based on the geometry of the network.
Figure~\ref{fig:voxel} shows a representation of the voxel tomography, with
rays crossing the voxels. At a specific time, one voxel can be crossed by many
rays, while other by very few rays. This fact and other
complexities of the tomography problem make it need regularization and
optimization. \citet{Haji-Aghajany2018} showed a method to solve some of these
issues, obtaining tomography values of water vapor density with high
spatio-temporal resolution, resulting in RMSE of \(0.39~\mathrm{g~m^{-3}}\). The
method was to use a ray-tracing technique in which rays are allowed to
move in the 3D space with a hybrid regularization method, which mixed direct
and iterative methods. The main idea was to project the large-scale problem into
a small-scale one that can be regularized by a direct method, and then transform
back to large-scale. The technique, apart from being more efficient
computationally, improved the results in some cases with respect to other
methods. A different method by \citet{Yang2019b} tried to maximize the number of rays
by considering water vapor density in the upper region outside the
boundaries as four (one per direction) new parameters to obtain. This addition
to the traditional method allowed the use of rays that do not come from the top
boundary of the study region, but from the sides. An experiment in Hong Kong was
carried out to test this method, showing better results than traditional
tomography. \citet{Yang2019b} also showed that, to improve tomography, it is more relevant to increase the number of voxels crossed by rays than than increasing the number of
rays.
However, the number of possible approaches is large and, therefore,
\citet{Brenot2019} compared several variations of approaches to tomographic techniques
with radiosonde and ECMWF reanalysis. This study recommended using stacked data
over a 30 min window (increasing the number of available rays) and pseudo-slant
observations in case studies and nowcasting. Regarding parametrization,
another study compared some methods using the European Cooperation in Science
and Technology (COST) data-set (central Europe) \citep{Adavi2020}. Among them,
ray-tracing was found to be more adequate than straight-line methods. In an
effort to increase the number of observations with respect to the number of
unknowns, \citet{Zhang2020a} proposed another method, which consisted on modeling the
height factor for both the isotropic and anisotropic part of slant wet delay
(SWD). This would allow signals that do not cross the tomography top boundary to
be included in the calculations, increasing the number of valid observations.

\begin{figure}[H]
\includegraphics[width=3.4in]{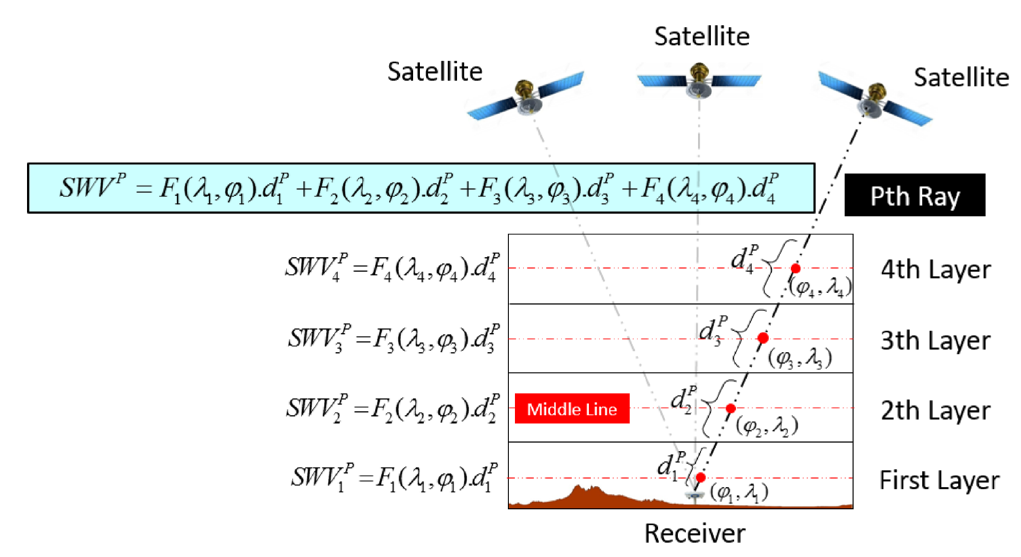} \caption{Schematics from \citet{Haji-Aghajany2020a} approach. Source: \citet{Haji-Aghajany2020a}.}\label{fig:haji2020}
\end{figure}

Voxel-based methodologies, despite being quite popular, have some drawbacks: (1)
the rank deficiency in the coefficient matrix, leading to the need of applying
constraints; (2) the assumption that water vapor is constant in the space
inside each voxel; and (3) the number of unknown parameters is high compared to
the number of observations. In order to avoid these drawbacks, some
methodologies avoiding the use of voxels have been developed. One of then considered a function-based
tomography, as Figure~\ref{fig:haji2020} shows. This approach uses B-spline
functions and a distribution of the atmosphere in layers as the base to build
the system of equations. A prior constraint of the 3D structure of the water
vapor needs to be applied in this vector, based on radiosonde data. A test
\citep{Haji-Aghajany2020a} with a 17-station network in California was carried out,
comparing the conventional, voxel-based method, and this new B-spline function
method, both validated against radiosonde observations. RMSE was reported to
reduce by about \(0.3~\mathrm{g~m^{-3}}\). Instead of B-spline functions,
\mbox{\citet{Zhang2020b} proposed} another method based on nodes rather than in voxels, with
the aim of reducing the discretization effects. Also, a piece-wise linear fitting
was used, instead of the standard (one-time) one. In an experiment to
compare the standard method and this node-based method against reference
radiosonde profiles, RMSE was found to decrease from \(1.5~\mathrm{g~m^{-3}}\) to
\(0.83~\mathrm{g~m^{-3}}\). The improvement was stronger in \mbox{rainy days}.

Whatever the tomography method proposed, another way to increase the number of
observations is to include several GNSS \citep{Dong2018}. The multi-GNSS tomography
approach was validated using the Continuously Operating Reference Stations
(CORS) network in Wuhan, China, against radiosonde and ERA5, improving in a
\(5\%\) GPS-only tomography. The comparison of GPS-only, GPS+GLONASS (Globalnaya
navigatsionnaya sputnikovaya sistema, Russian GNSS) and GPS+GLONASS+BDS (BeiDou
Navigation Satellite System, Chinese GNSS), showed that BDS did not improve much
the tomography. The improvements were associated to the increase of signals
available for tomography (\(1.6\) times more for 2-GNSS, and \(2.2\) for the three
GNSS).

Tomography of water vapor using GNSS networks is an emerging field that,
although it has room for improvement, will provide very useful information
about the 3D water vapor distribution. The use of multi-GNSS approaches and
the densification of station networks will help improving tomography. This
will improve NWP analysis and forecasts, and the general knowledge of
atmospheric water vapor.

\hypertarget{combining-with-other-measurements.}{%
\subsection{Combining with Other Measurements.}\label{combining-with-other-measurements.}}

Being the atmospheric water vapor highly varying in space and time, the
combination of different data-sets with different characteristics can enhance
the knowledge of its spatial and temporal distribution. For example,
\citet{Furumoto2003} proposed an algorithm to use GNSS-derived water vapor measurements
to improve humidity profiles estimated from middle and upper atmosphere radar
with the radio acoustic sounding technique.

The use of satellite measurements with GNSS data is an interesting idea,
since satellite instruments have opposed characteristics to GNSS: high
spatial, but low temporal resolution. They are also generally not available for
all weather situations, but GNSS data are available for all weather situations without important
quality loss. One example of this approach \citep{Lindenbergh2008} used MEdium
Resolution Imaging Spectrometer (MERIS) satellite observations combined with GNSS
measurements to estimate IWV by means of a kriging methodology.
Thus, IWV was retrieved at a time and point in space by means
of (1) a kriging combination of GNSS IWV from several locations at that same time; and
(2) MERIS IWV from that point in space at the closest time available. Another
example of combination of satellite and GNSS IWV was provided by \citet{Leontiev2018a}.
This approach used Meteosat-10 data to identify cloud features, and then using
this information in the GNSS IWV interpolation (Delaunay triangulation or kriging
interpolation) process. Using this approach in the Israel area, and validating
against radiosonde, the obtained bias and RMSE were \(1.77\) and \(2.81~\mathrm{kg~m^{-2}}\) to
\(0.74\) and \(2.04~\mathrm{kg~m^{-2}}\), respectively.

Other works combine GNSS data with models. An experiment ingested GNSS and
Interferometric synthetic-aperture radar (InSAR) IWV and the Weather Research
and Forcasting (WRF) model to cover areas that are underrepresented by GNSS/InSAR
stations, using a fixed-rank kriging approach \citep{Alshawaf2015}. The results were
compared to MERIS imagery, and uncertainties of less than \(1~\mathrm{mm}\) were
found. One of the strengths of this technique is that more data sources can be
used, improving further the quality of the generated IWV maps. Another hybrid
data-fusion model has been proposed, using GPT2 model as base, and correcting it
with polynomial fitting and spherical harmonics functions obtained from GNSS and
radiosonde IWV, also considering correction for height differences associated to
each station \citep{Zhao2020}.

Data fusion approaches allow increasing even more the benefits of the
continuously operating GNSS networks, which have the potential to enhance
weather forecasts and analysis. When applying them, it is important to correct
the IWV for height differences, since IWV decrease with increasing height.

\hypertarget{ic}{%
\section{Inter-Comparisons with Other Techniques}\label{ic}}

\hypertarget{validations-of-gnss-water-vapor-with-other-techniques-as-reference}{%
\subsection{Validations of GNSS Water Vapor with Other Techniques as Reference}\label{validations-of-gnss-water-vapor-with-other-techniques-as-reference}}

Table~\ref{tab:ic-gnss} presents a summary of the main findings of the works used to develop this subsection.
One of the first GNSS IWV validations as an emerging system for water vapor
monitoring was performed by \citet{Duan1996a}. They applied an approach that
incorporated GNSS stations that were far away (long baselines) to estimate the
absolute ZTD, while previous methodologies used WVR for calibration of the
relative ZTD path. This could lead to errors, since WVR overestimate IWV during
rainfall. Then, they compared IWV at a few stations in Oklahoma and Kansas with
nearby WVR stations, obtaining negligible bias and RMSE in the interval
1.0--1.5~\(\mathrm{mm}\). It was proved that GNSS could, by itself, provide
accurate ZTD values without the need of auxiliary instruments, which was
specially important for all-weather IWV retrieval, since WVR typically fails
under intense rainfall. Another work analyzed the usefulness of GNSS for IWV
monitoring, and compared the estimates against NWP. It was found that, although
GNSS IWV showed the potential to become a powerful tool for improving forecasts, the
products still needed to be refined, as biases were found in the
2--4~\(\mathrm{mm}\) and RMSE of \(3.8~\mathrm{mm}\) \citep{Iwabuchi2000}. In this work,
it must be noted that the meteorological data used for converting ZTD to IWV was
obtained from the NWP itself, and therefore not as accurate as a meteorological
station, specially for near real-time applications. There were also validations
against radiosonde stations, which are usually a reference for water vapor
sensing. For instance, \citet{Basili2001} studied the comparison of GNSS and WVR against
radiosondes during 1999 in non-precipitating conditions, using the Cagliari
(Italy) GNSS station and nearby radiosonde and WVR sites. The bias between GNSS
and radiosondes was \(0.4~\mathrm{mm}\) and the standard deviation
\(1.9~\mathrm{mm}\), which already showed better results than \citet{Iwabuchi2000}.

It is useful to test the slant-IWV, as it can be used in some applications
(for instance, tomography) and it is also an interesting exercise to evaluate GNSS
observations without the possible errors from mapping functions. \citet{Braun2003}
compared line-of-sight integrals of water vapor from GNSS using WVR as a reference
in Oklahoma. They found very good agreement between both techniques
(correlation \(0.99\), RMSE \(1.3~\mathrm{mm}\)). This work concluded that GNSS data
provided an all-weather alternative that is especially useful under rainfall or sever
weather events, situations in which other instruments (such as WVR itself) fail.

\newpage
\end{paracol}
\nointerlineskip
\begin{specialtable}[H]
\widetable
\caption{\label{tab:ic-gnss}Summary of studies on validation of GNSS measurements. Column Year is the publication year. N is the number of stations. Instr. is the instruments (besides ground-based GNSS) used in the study. Start and End mark the period considered in the study, in year-month format. Bias is the bias in mm, and DEV is the RMSE or SD also in mm. Corr is the correlation coefficient (R).}

\begin{tabular}{m{0.8cm}<{\centering}m{1cm}<{\centering}m{0.6cm}<{\centering}m{2cm}<{\centering}m{1.2cm}<{\centering}m{1.2cm}<{\centering}m{2cm}<{\centering}m{1.8cm}<{\centering}m{1.8cm}<{\centering}m{1.5cm}<{\centering}}
\toprule
\textbf{Work} & \textbf{Year} & \textbf{N} & \textbf{Instr.} & \textbf{Start} & \textbf{End} & \textbf{Region} & \textbf{Bias} & \textbf{DEV} & \textbf{Corr}\\
\midrule
\citep{Duan1996a} & 1996 & 4 & WVR & \hl{1993-05} %MDPI: please check if this is correct.
 & \hl{1993-05} &  {Interior US  (US)} & 0.05 & 1.28 & -\\
\midrule
\citep{Iwabuchi2000} & 2000 & 556 & NWP & \hl{1996-09} & \hl{1996-09} & Japan & 2--4 & 3.8 & -\\
\midrule
\citep{Basili2001} & 2001 & 1 &{radiosonde, WVR} & \hl{1999-01} & \hl{1999-12} &  {Cagliari  (Italy)} & 0.4 & 1.9 & -\\
\midrule
\citep{Braun2003} & 2003 &  & WVR & \hl{2000-05} & \hl{2000-06} & Oklahoma & - & 1.3 & 0.99\\
\cmidrule{1-2}
\cmidrule{4-10}
\citep{Memmo2005} & 2005 & \multirow[t]{-2}{*}{\raggedright\arraybackslash 5} &{MWR, NWP (MM5), radiosonde} & \hl{2002-08} & \hl{2003-09} &  {Europe  (Italy)} & 0.01 & 0.286 & 0.939\\
\midrule
\citep{Glowacki2006} & 2006 & 8 &  {NWP(GASP; ECMWF), radiosonde} & \hl{2000-01} & \hl{2000-12} & Australia &  {$-$1.4--+1.1 (GASP), $-$1.8--+1.3 (ECMWF)} &{3.1--5.3 (GASP), 2.7--3.5 (ECMWF)} & -\\
\midrule
\citep{Ha2010a} & 2010 & 3 & {MWR, radiosonde} & \hl{2009-08} & \hl{2009-09} &  {Ulleungdo  (South Korea)} & {:3.0 (Radiosonde), 5.0 (MWR)} & {12.5 (Radiosonde), 9.3 (MWR)} & -\\
\midrule
\citep{Musa2011} &  & 4 & radiosonde & \hl{2008-01} & \hl{2008-12} & Malaysia & $-$1.5--0.0 & 3.5--4.3 & 0.80--0.88\\
\cmidrule{1-1}
\cmidrule{3-10}
\citep{Thomas2011c} & \multirow[t]{-2}{*}{\raggedleft\arraybackslash 2011} & 12 &{radiosonde, AIRS, MODIS} & \hl{2004-01} & \hl{2004-12} & Antarctic & $-$0.48 (Radiosonde) & - & -\\
\midrule
\citep{Vaquero-Martinez2019h} & 2019 & 4 & radiosonde & \hl{2006-05} & \hl{2018-01} & global & $-$0.87--($-$0.49) & 0.61--1.10 & 0.99\\
\bottomrule
\end{tabular}
\end{specialtable}
\begin{paracol}{2}
%\linenumbers
\switchcolumn

The potential of GNSS IWV for model verification was assessed \mbox{by \citet{Guerova2005}}.
GNSS was starting to be acknowledged as a reliable technique for validating models
and other instruments. Particularly, this work includes some validation of GNSS
IWV against radiosondes and MWR in Switzerland between 2001 and June
2003. During daytime, GNSS showed a positive bias against radiosonde
(\(0.9~\mathrm{mm}\)), while at night, the bias became negative
(\(-0.4~\mathrm{mm}\)). However, against MWR, both night and daytime GNSS IWV
showed positive bias (\(0.3~\mathrm{mm}\)). Therefore, in general terms, it was
concluded that GNSS IWV was an accurate technique, and it was used to evaluate
Alpine Model analysis and \mbox{forecast \citep{Guerova2005}}.
The model showed a seasonal
dependence with an increased dry bias in summer 2002. The main conclusion was
that GNSS provide information accurate enough for model verification and
assimilation. Another work \citep{Memmo2005} compared the Fifth-Generation Penn
State/NCAR Mesoscale Model (MM5) forecast IWV against GNSS, radiosonde and
microwave radiometer (MWR), showing that summer had worse performance (higher
errors, smaller correlation) than other seasons. In this work, summer also had a
higher bias (wet in this case) but this is associated to the fact that IWV
values are higher in summer. The diurnal cycle is also reported to be an aspect
to improve, and errors are mainly associated to errors in the temperature daily
cycle by MM5. Therefore, in this work, GNSS IWV was successfully used as a
reference to find deficiencies in models.

However, one of the best tests for GNSS IWV products is to compare them against
radiosonde measurements, a direct method usually considered a reference despite
known bias issues, see \citep{Dirksen2014}.
\citet{Glowacki2006} compared GNSS IWV from eight stations in Australia against
collocated or nearby radiosonde stations, obtaining standard deviations of
\(8.8\%\), except for Antarctic stations, which increased to 18--21\(\%\). It
must be noticed that GNSS is considered less sensitive to small amounts of
water vapor and, therefore, in regions with small IWV, GNSS technique can perform
worse. This was assessed by a cross-validation including, FTIR, sun-photometer,
multifilter rotating shadowband radiometer (MFRSR), radiosonde, and GNSS in the
Izaña Observatory (Canary Islands, Spain) during years 2005--2009, showed similar
biases with respect to GNSS (\(-3.58\) to \(+0.42~\mathrm{mm}\)), and standard deviations
(between \(0.73\) and \(2.29~\mathrm{mm}\)), which is consistent with other works.
GNSS accuracy was found to be better than \(10\%\) for values over
\(3.5~\mathrm{mm}\). However, for values under this limit, the accuracy is worse (\(\sim\)20\%) and a systematic dry bias appears \citep{Schneider2010a}.

Other comparisons against radiosondes, like the one set in four
stations of the Malaysian Peninsula by
\citet{Musa2011} during the year 2008, or the one with four collocated reference GRUAN
radiosonde stations \citep{Vaquero-Martinez2019h}, assessed the high quality of
GNSS to monitor water vapor. RMSE values were in the range
3.5--4.3~\(\mathrm{mm}\) in the former, while standard deviations were
0.61--1.10~\(\mathrm{mm}\) in
the latter, and correlations were of  0.80--0.88, reaching \(0.99\) in the
second work. Both works presented dry biases, varying between \(0.0\) and
\(-1.5~\mathrm{mm}\) in the first cases, and 0.49--0.87~\(\mathrm{mm}\) in the
second. The different biases are mainly associated to dry bias in the
radiosonde measurements, while the better results in the second work could be
due to the use of GRUAN stations, with high quality standards, and also to
the fact that both radiosondes and GNSS stations were exactly collocated.
These works show that GNSS data are high quality but, in any case, the
presence of redundant measurements help assess the quality of products.
Radiosonde dry bias was also assessed by \citet{Bock2007}.
This work inter-compared GNSS, radiosonde, Special Sensor Microwave/Imager (SSM/I) satellite instrument (in coastal
stations only) and NWP (ERA-40 and National Centers for Environmental Prediction 2, NCEP2) for stations in Africa. The standard
deviations were of \(3.0\), \(2.1\),\(3.6\), and \(5.2~\mathrm{mm}\), respectively,
while the biases (GNSS - instrument) were \(3.2\), \(-0.6\), \(0.5\), and
\(1.6~\mathrm{mm}\), respectively. The results confirmed the high potential of GNSS
for water vapor monitoring. Another work found a cause for dry bias in tropical
regions \citep{Zhang2018d}: the Integrated Global Radiosonde Archive (IGRA) product
only integrates water vapor up to a pressure level of \(500~\mathrm{hPa}\). While
this is usually a fair approximation, in tropical regions it induces a
dry bias in the radiosonde product; in the case of Tahiti, of 9\%.

Another interesting region for validation is Antarctica. It is a region with relatively scarce water vapor observation data; yet its water vapor spatial distribution and temporal variability influences precipitation patterns and, therefore, ice accumulation. Hence, a GNSS-derived data set for Antarctica, with proven accuracy and stability, could be very useful for modeling and obtaining trends \citep{Thomas2011c}. A work used
homogeneously reprocessed GNSS data for the year 2004 in 12 stations
\citep{Thomas2011c} in this region. These data had three important improvements
compared to previous GNSS IWV products: first, it used absolute antenna phase
center variations. Second, it used VMF1 mapping functions. And third, an
accurate a priori model for ZHD. For comparison, radiosonde measurements were
used as reference, finding a GNSS bias of \(-0.48~\mathrm{mm}\), very adequate even
for a region with low IWV that could affect GNSS sensitivity to water vapor.
Atmospheric Infrared sounder (AIRS) and Moderate Resolution Imaging
Spectroradiometer (MODIS) IWV products were also compared to GNSS in this work,
with low bias (\(0.58\) and \(0.35~\mathrm{mm}\), respectively) and RMSE
(\(1.24~\mathrm{mm}\) and \(1.42~\mathrm{mm}\)). This concluded that the newly
analyzed GNSS IWV product had enhanced capabilities for assimilation in NWP.

Multi-comparisons can help attribute errors to
one or another instrument, making easier to analyze deeper the possible causes
of error. As an example, a work that compared radiosonde, GNSS,
Fourier-transform infrared spectroscopy (FTIR), MWR, Advanced Microwave Sounding
Unit (AMSU-B) satellite based microwave radiometer, and ERA-Interim NWP in the
location of Kiruna (Sweden) \citep{Buehler2012a} found systematic differences in the
range of \(\pm 1~\mathrm{mm}\), consistent with other mentioned works. Although
the overall comparison shows a good agreement among the different techniques,
they aimed to characterize the slope and offset of the different techniques,
combining their study with others, but no consistent pattern emerged (the only
pattern was that systematic differences were in the \mbox{1--2~\(\mathrm{mm}\)} range). This
work also presented a method to compare values of IWV with different heights
(that is, where the lower limit of the integral starts a different heights),
based on a correction factor linearly dependent with height. The factor was
obtained with radiosonde data. The representativeness error was also addressed,
presenting a method to estimate it and, finding that it can dominate the
differences among techniques, what could make it so that observed differences do not
contain much information about the true precision of the techniques. This must
be taken into account in future works to make validations more statistically
significant.
Another interesting inter-comparison \citep{VanMalderen2014b} involved radiosondes,
GNSS, sun-photometers, a combination of GOME, GOME-2 and SCIAMACHY products, and
AIRS IWV product. The study was carried out in 28 sites in Northern
Hemisphere. It was found that biases were in the \(-0.3\) to \(+0.5~\mathrm{mm}\)
interval, with larger standard deviation (SD) values (0.96--3.96~\(\mathrm{mm}\)) specially for
satellite instruments. Correlations with GNSS IWV were larger for ground-based
instruments (\(R^2 > 0.97\)) while slightly smaller for satellite instruments
(\(R^2 > 0.88\)). Sun-photometers and satellite retrievals were shown to need low
cloud fractions to perform properly, being this more important than proximity
between GNSS station and satellite pixel center or sun-photometer station. All
instruments showed worse performance with respect to GNSS as cloud fraction
increases, which in the case of satellite instruments can be explained for the
so-called screening effect. As cloudy skies typically have larger IWV, the
instruments show stronger dry bias for high IWV. Radiosonde and AIRS measure
both in daytime and nighttime, and both showed a wet bias during nighttime,
being reduced (radiosonde) or changed to dry bias (AIRS) during daytime. In any
case, inter-comparisons are reported to be site- and instrument-dependent, as
\citet{Buehler2012a} indicated.

Not only IWV has been tested, but also profiles obtained through tomography.
\mbox{\citet{Ha2010a} validated GNSS} water vapor profiles with collocated MWR and radiosonde
observations, using MWR as a reference in Ulleungdo (Korea) for a few days in July
2009. It was found that GNSS tomography could even improve radiosonde's RMSE
values, but in some cases GNSS profiles were degraded, which was associated with
the fact that only \mbox{two GNSS} stations were used in the tomography. Under severe
weather, GNSS showed very good agreement with MWR, in comparison with radiosonde.
This work concluded that another important issue for improving GNSS tomography was
the satellite geometry, which is expected to improve with multi-GNSS approaches.

The different validations show that GNSS has great potential as a tool for IWV
monitoring. It is important to put an effort into homogenizing the series, especially
when calculating trends and, when performing comparisons, taking into account
possible height differences and other deficiencies in the instruments considered.
These efforts help reduce spurious biases and variability. The GNSS
technique has the advantage of being relatively cheap, weather-independent and with high
temporal resolution, and for these reasons this technique is preferred to
others in many cases.

\hypertarget{validations-of-nwp-with-gnss-water-vapor-as-reference}{%
\subsection{Validations of NWP with GNSS Water Vapor as Reference.}\label{validations-of-nwp-with-gnss-water-vapor-as-reference.}}

Table~\ref{tab:ic-nwp} presents a summary of the main findings of the works used to develop this subsection.
GNSS ground-based stations have been widely used for testing NWP. One of the
first works of this kind validated High Resolution Limited Area Model (HIRLAM)
against GNSS IWV in Spanish (Madrid Sierra) station \citep{Cucurull2000}, obtaining
a bias of \(-0.4~\mathrm{mm}\) and RMSE of \(2.0~\mathrm{mm}\) during two short
periods in December 1996. Another work \mbox{by \citet{Kopken2001a}} compared the Baltic Sea Experiment
(BALTEX) model and the Europa model against \mbox{25 GNSS} stations in Finland and
Sweden. They found relatively high biases (2.5--3.0~\(\mathrm{mm}\)), and adequate
standard deviations (\(2.8~\mathrm{mm}\) on average) and correlations (\(0.92\) on
average), but with worse results than HIRLAM.
Testing against radiosondes yielded smaller biases, what suggested GNSS
could have a slight dry bias in that region.
SSM/I was also analyzed in this work and showed a good correlation with
the GNSS product, assessing the use of both independent instruments for validation
and ingestion in analysis.
NCEP reanalysis was deeply
validated at a global scale against GNSS by \citet{Vey2010a}, for a ten-year
period (1994--2004) at daily resolution. The study considered 11 clusters of
stations for regional analysis. The clusters were based on humidity regimes and
geographical information. Also, homogeneity of time-series from GNSS stations was
analyzed to select only homogeneous time-series or to correct those which had jumps in
the series. The validation of NCEP covered several aspects: correlation, bias,
seasonal signal, anomalies and variability. Correlation was shown to be very
close to the unit, in the range 0.89--0.99, except in some regions of South America
and the Antarctic Peninsula, which are regions that have a small number of
observations available to be assimilated by the model and, therefore,
worse performance is expected. Regarding bias, most stations were in the \(\pm 3.0~\mathrm{mm}\) range, and the biases were associated to differences between the
model height and the GNSS station height (which causes errors due to the
assumption of temperature lapse rate). The biases were also associated with
stations with high elevation, where the reduction of pressure to station level
could also be error prone. Biases in stations that did not fit with these two
problems (i.e.,~large elevation differences or large elevations) are associated
with NCEP problems, such as inconsistencies in assimilated observations and inability
of the model to resolve local changes. Regarding the seasonal signal, it is found to
be strongly underestimated in Antarctica, while having a good agreement in
Europe and most of North America. The computation of anomalies suggest
sub-millimeter accuracy in GNSS, assessing its suitability for climate studies,
providing that the processing strategy remains unchanged and modifications of
hardware or number of GNSS observations were considered in the analysis. The study
also concluded that the Southern Hemisphere needs more observations for reanalysis
assimilation, suggesting the use of GNSS IWV to accomplish this.

However, not only direct comparisons are performed, but also other aspects, like
the effect of different spatial resolution in NWP, are studied. \citet{Yang1999}
evaluated HIRLAM against GNSS network in Northern Europe, along with some
forecasts. It was found that relatively coarse resolution (\(0.4^\circ\)) runs had
a similar performance to finer resolution (\(0.21^\circ\)) ones. Therefore, a finer
resolution did not mean better agreement with observations. It was also found
that NWP are useful for detecting erroneous data in GNSS series.

Other studies have focused on the causes of discrepancies between NWP and GNSS.
The height difference between the orography of the model and the GNSS station can
induce systematic differences, since the integrals are
not measured from the same base level. This issue was studied by \citet{Matsuyama2020},
who compared Japanese 55-year Reanalysis (JRA-55) and GNSS in 26 Japanese
stations in the period July 2010 to December 2012. After JRA-55 was corrected to
account for the height difference between the GNSS station and the JRA-55 integral base
level of geopotential, some dry bias in JRA-55 IWV still remained. This
suggested that not all of the discrepancy in IWV could be attributed to the height
difference and, therefore, the hydrological cycle of this reanalysis needed
deeper analysis. Moreover, NWP can provide inaccurate data when the observations
assimilated have erroneous or biased data. For example, some radiosonde models
have a well-known issue that caused a dry bias. To study this problem,
\citet{Bock2009} compared radiosondes and NWP (ECMWF Integrated Forecasting System, and
NCEP reanalysis I and II) against GNSS IWV in
western Africa over 2005--2008. In this case, the Vaisala RS80-A radiosonde had a dry bias
leading to errors in NWP through assimilation,
causing a dry bias of NWP with respect to GNSS. Other two radiosonde models
(Vaisala RS92 and MODEM) seemed to cause a wet bias in NWP models. Another issue is how a pixel from NWP is
representative of a point (i.e., a GNSS station) in the area it covers.
\citet{Bock2019} addressed the representativeness error that ERA-Interim can have
with respect to GNSS in stations distributed globally. A ``representativeness
error statistic'' was developed, showing a correlation of \(0.73\) with standard
deviation of IWV differences between ERA-Interim and GNSS. This statistic is
based on the information contained in the ERA-Interim data-set for the four
surrounding pixels for the GNSS station. To reduce the representativeness
error, it was recommended to bilinearly interpolate data to the position of the
GNSS station, as well as time-averaging. Some problematic cases (stations whose
data do not fit as well as the rest) were discussed.

\end{paracol}
\nointerlineskip
\begin{specialtable}[H]
\widetable
\caption{\label{tab:ic-nwp}Summary of studies on validation of NWP data. Column Year is the publication year. N
is the number of stations. Instr. is the instruments (besides ground-based GNSS) used in the study.
Start and End mark the period considered in the study, in year-month format. Bias is the bias in
mm, and DEV is the RMSE or SD also in mm. Corr is the correlation coefficient (R).}

\begin{tabular}{m{0.8cm}<{\centering}m{1cm}<{\centering}m{0.6cm}<{\centering}m{2.5cm}<{\centering}m{1.2cm}<{\centering}m{1.2cm}<{\centering}m{2.8cm}<{\centering}m{2.4cm}<{\centering}m{0.7cm}<{\centering}m{0.7cm}<{\centering}}
\toprule
\textbf{Work} & \textbf{Year} & \textbf{N} & \textbf{Instr.} & \textbf{Start} & \textbf{End} & \textbf{Region} & \textbf{Bias} & \textbf{DEV} & \textbf{Corr}\\
\midrule
\citep{Yang1999} & 1999 & 25 & HIRLAM & \hl{1995-08} %MDPI: plesae check if thisis correct. is this year and month?
 & \hl{1995-11} & {Northern Europe (Finland, Sweden)} & $-$0.1 & 2.3 & 0.94\\
\midrule
\citep{Cucurull2000} & 2000 & 5 & {HIRLAM, radiosonde} & 1996-12 & 1996-12 &  {Madrid Sierra  (Spain)} & $-$0.4 & 2.0 & -\\
\midrule
\citep{Kopken2001a} & 2001 & 25 & {BALTEX; European, SSM/I, WVR} & 1995-08 & 1995-10 &  {Finland, Sweden} & 2.5--3.0 & 2.8 & 0.92\\
\cmidrule{1-10}
\citep{Bengtsson2004} & 2004 & 160 &  {ERA-40; NOSAT} & 2000-07 & 2001-01 & global & $-$31.26--10.75 \% & - & -\\
\midrule
\citep{Guerova2005} & 2005 & 20 & {MWR, Alpine Model, radiosonde} & 2001-01 & 2003-06 & Switzerland & {$-$2.8 (forecast), $-$1.5 (analysis)} & - & -\\
\midrule
\citep{Bock2009} & 2009 & 11 & {ECMWF, NCEP, radiosonde} & 2005-01 & 2008-12 & Africa & - & - & -\\
\midrule
\citep{Parracho2018a} & 2018 & 104 &  &  & 2010-12 & global & - & - & -\\
\cmidrule{1-3}
\cmidrule{6-10}
\citep{Bock2019} & 2019 & 120 & \multirow[t]{-2}{*}{\raggedright\arraybackslash ERA-Interim} & \multirow[t]{-2}{*}{\raggedright\arraybackslash 1995-01} & 2010-12 & global & $-$1--1 & <2 & -\\
\midrule
\citep{Matsuyama2020} & 2020 & 26 & JRA-55 & 2010-07 & 2012-12 & Japan & - & - & -\\
\bottomrule
\end{tabular}
\end{specialtable}
\begin{paracol}{2}
%\linenumbers
\switchcolumn

Nevertheless, not only is IWV tested in NWP, but also its trends. Trends of IWV
are very important for discussing how climate is evolving over the years. For
this reason, \citet{Bengtsson2004} discussed whether reanalysis data can provide valid trends for
IWV and other variables. Particularly, IWV and its trends, computed with the ERA40
and NOSAT experiment (reanalysis without satellite data assimilation to mimic
the pre-satellite era), were compared globally against GNSS data. ERA40 trends
were corrected to avoid spurious trends due to changes in the observation
system. In any case, the trends could not be properly reproduced when compared
to GNSS data. Later, \citet{Parracho2018a} studied trends from ERA-Interim and
Modern-Era Retrospective analysis for Research and Applications, Version 2
(MERRA-2) IWV data-sets for the period 1980--2016, as well as for 104 GNSS sites
globally from 1995 to 2010. It was found that ERA-Interim was generally
consistent with GNSS trends, while MERRA-2 tended globally to more moistening
trends. Inconsistent trends between both reanalyses were found in Antarctica
and most of the Southern Hemisphere, as well as central and Northern Africa.
These regions have a high uncertainty in both reanalysis, due to the low
number of in-situ observations, and the dispersion between one model and the
other is generally high. Arid regions are found to have an important dependence
of water vapor on circulation patterns, and therefore, in those regions the
Clausius-Clapeyron scaling ratio is not a good proxy to infer IWV inter-annual
variability and decadal trends.

GNSS IWV products have proven to be a reliable source of data for testing NWP
performance. However, some issues must be addressed, like the homogenization of
time-series (specially when dealing with trend estimation) and possible height
differences between GNSS station and pixel in the orography of the NWP.

\hypertarget{validation-of-satellite-measurements-with-gnss-as-reference.}{%
\subsection{Validation of Satellite Measurements with GNSS as Reference.}\label{validation-of-satellite-measurements-with-gnss-as-reference.}}

Table~\ref{tab:ic-sat} presents a summary of the main findings of the works used to develop this subsection.
GNSS IWV data have been widely used as a reference to validate satellite IWV
products. Satellite radiometers use different regions of the electromagnetic
spectrum in which water vapor have absorption bands. One of the most common is
the infra-red (IR) or near infra-red (NIR), in which water vapor is the main absorbent when compared to the
rest of atmospheric species. However, liquid water masses, like clouds, can
disrupt the signal. This is the so-called screening effect: water vapor below
the cloud remains ``hidden'' to the satellite radiometer, resulting in a dry bias.
This is not always reflected in the comparisons, probably due to corrections
in the retrieval algorithms.
\end{paracol}
\nointerlineskip
\begin{specialtable}[H]
\widetable
\caption{\label{tab:ic-sat}Summary of studies on validation of satellite measurements. Column Year is the publication
year. N is the number of stations. Instr. is the instruments (besides ground-based GNSS) used
in the study. Start and End mark the period considered in the study, in year-month format. Bias is
the bias in mm, and DEV is the RMSE or SD also in mm. Corr is the correlation coefficient (R).}
\scalebox{0.91}[0.91]{
\begin{tabular}{m{0.8cm}<{\centering}m{1cm}<{\centering}m{0.6cm}<{\centering}m{3cm}<{\centering}m{1.2cm}<{\centering}m{1.2cm}<{\centering}m{2.8cm}<{\centering}m{1.8cm}<{\centering}m{1.5cm}<{\centering}m{1.5cm}<{\centering}}
\toprule
\textbf{Work} & \textbf{Year} &\textbf{N} & \textbf{Instr.} & \textbf{Start} & \textbf{End} & \textbf{Region} & \textbf{Bias} & \textbf{DEV} & \textbf{Corr}\\
\midrule
\citep{Raja2008} & 2008 & 375 & AIRS & \hl{2004-04} %MDPI: please check if this is year and month.
 & 2004-10 & US & 0.5--1.2 & 3.0--4.5 & 0.91--0.98\\
\midrule
\citep{Bennouna2013b} & 2013 & 18 & MODIS & 2002-01 & 2008-12 &  {Iberian P. (Spain)} & 0.1--0.4 & 4.4--5.7 & 0.83--0.88\\
\midrule
\citep{VanMalderen2014b} & 2014 & 28 & {radiosonde, AIRS, GOME, GOME-2, SCIAMACHY, sunphotometer} & 1995-01 & 2011-04 & Northern Hemisphere & - & - & -\\
\midrule
\citep{Roman2015b} & 2015 & 21 & GOME2 & 2007-01 & 2012-12 &  {Iberian P.  (Spain)} & 0.7 & 4.4 & 0.84\\
\midrule
\citep{Nelson2016} &  & - & {MWR, radiosonde, OCO-2, sunphotometer} & 2014-09 & 2016-02 & global & <0.5 & 1.3 & 0.994\\
\cmidrule{1-1}
\cmidrule{3-10}
\citep{Ningombam2016a} &  & 1 & MODIS-NIR &  & 2012-12 & Trans-Himalayan & $-$0.18 & 1.37 & 0.9848858\\
\cmidrule{1-1}
\cmidrule{3-4}
\cmidrule{6-10}
\citep{Wang2016d} & \multirow[t]{-3}{*}{\raggedleft\arraybackslash 2016} & 250 &  {OMI, SSMIS, sunphotometer} & \multirow[t]{-2}{*}{\raggedright\arraybackslash 2005-01} &  & global & - & - & -\\
\cmidrule{1-5}
\cmidrule{7-10}
\citep{Vaquero-Martinez2017c} &  & 9 & OMI &  & \multirow[t]{-2}{*}{\raggedright\arraybackslash 2009-12} &  {Iberian P.  (Spain)} & $-$0.3 & 5.1(IQR) & 0.79\\
\cmidrule{1-1}
\cmidrule{3-4}
\cmidrule{6-10}
\citep{Vaquero-Martinez2017d} & \multirow[t]{-2}{*}{\raggedleft\arraybackslash 2017} & 21 & MODIS &  &  & {Iberian P.  (Spain)} & 0.9\% & 39.36\% (IQR) & 0.78\\
\cmidrule{1-4}
\cmidrule{7-10}
\citep{Vaquero-Martinez2018c} & 2018 & 9 &  {AIRS, GOME-2, MODIS, OMI, SCIAMACHY, SEVIRI} & \multirow[t]{-3}{*}{\raggedright\arraybackslash 2007-01} & \multirow[t]{-2}{*}{\raggedright\arraybackslash 2012-12} &  {Iberian P.  (Spain)} & $-$5.2--16.7\% & - & 0.75--0.91\\
\midrule
\citep{Gong2019} &  & 1 & {FengYun-3A-MERSI, MODIS, sunphotometer} &  & 2013-07 & Tibetian Plateau & $-$1.14--0.64 & 1.87--2.75 & -\\
\cmidrule{1-1}
\cmidrule{3-4}
\cmidrule{6-10}
\citep{He2019} &  & 370 & {ENVISAT-MERIS, FengYun-3A-MERSI, Terra-MODIS} & \multirow[t]{-2}{*}{\raggedright\arraybackslash 2010-01} & 2010-12 & US & $-$5.9--5.6 & 2.5--12.4 & 0.73--0.94\\
\cmidrule{1-1}
\cmidrule{3-10}
\citep{Wang2019e} & \multirow[t]{-3}{*}{\raggedleft\arraybackslash 2019} & 303 &{OMI, SSMIS} & 2006-01 & 2006-12 & global & $-$3--4 & 3--10 & -\\
\midrule
\citep{CarbajalHenken2020a} &  & 300 &  {IASI, MIRS, MODIS, MODIS-FUB} & 2012-05 & 2014-06 &  {Europe  (Germany)} & $-$1.7--1.4 & 2.52--3.77 & 0.88--0.96\\
\cmidrule{1-1}
\cmidrule{3-10}
\citep{Vaquero-Martinez2020c} & \multirow[t]{-2}{*}{\raggedleft\arraybackslash 2020} & 1 & {AIRS, GOME2, MODIS, OMI, POLDER, SCIAMACHY} & 2010-01 & 2017-12 & European Arctic & $-$2.67--5.87 & 2.36--7.32 & 0.47--0.85\\
\bottomrule
\end{tabular}}
\end{specialtable}
\begin{paracol}{2}
%\linenumbers
\switchcolumn
Another common region for water vapor sensing is the visible, typically red or
blue bands. In this case, we have the same problem as with infra-red: clouds are
opaque to visible light. In addition, water vapor bands are not so strong in this
region, and therefore the sensitivity will not be as high as in the IR
case. However, this can be an advantage in some situations. When there is a high amount of
IWV, the infra-red absorption can saturate and therefore we would not be able to
sense water vapor over a certain threshold. Nevertheless, in that case, the
visible bands would not saturate.

The microwave, which GNSS also use, is another common region for extracting
information about water vapor. This region has the advantage of being more
independent of clouds. However, passive systems (that is to say, radiometers that
just receive the electromagnetic radiation reflected by the Earth's surface and
atmosphere) receive very small amounts of microwave energy. This makes it
difficult to properly sense water vapor with this kind \mbox{of instrument}.

As we show in the following lines, different works provide very different values
depending mostly on three parameters: the region studied, the specific satellite
product and the quality control applied. Other sources of differences can be the
use of averaged data (daily or monthly means, for example) and the period
involved.

One of the first validations of satellite IWV products using GNSS as a
reference was carried out by \citet{Raja2008}. They compared AIRS IWV retrieval against
GNSS data from a network in the United States (375 stations) during April--October 2004. Bias (AIRS-GNSS)
were in the range 0.5--1.2~\(\mathrm{mm}\), while RMSE values were between
\(3.0\) and \(4.5~\mathrm{mm}\). Correlations (0.91--0.98) were calculated monthly. The
biases were found to have a seasonal behavior (from negative in April, to peak
positive in July and then decreasing). It was concluded that GNSS data are an
excellent tool for satellite validation, because they allow all-weather
verification, with a high number of pairs thanks to their high temporal resolution
(30 min in this work), and their accuracy. The quality of AIRS
retrievals was assessed.

The Iberian Peninsula has been widely tested, in part because of two main
reasons: first, it is a region that gives home to different climate regimes in a
relatively small area and, second, it is a region with a relatively dense
network of GNSS stations and collocated (or nearby) meteorological stations.
\citet{Bennouna2013b} studied the performance of MODIS-IR and NIR IWV annual cycle
against ground-based techniques, mainly GNSS, in 18 locations in the Iberian
Peninsula. The comparison is very detailed by zones of this region and by
season, finding high differences in MODIS performance between coastal and
interior stations. Representativeness error is also studied, finding that using
only time-coincident data for monthly means improves the agreement between MODIS
and GNSS. RMSE were found to be between \(4.4\) and \(5.7~\mathrm{mm}\), biases of
0.1--0.4~\(\mathrm{mm}\), and correlations of 0.83--0.88. Global Ozone Monitoring
Experiment 2 (GOME-2) satellite instrument was later evaluated by \citet{Roman2015b} in
this region, using 21 stations during the years 2007--2012. Bias was found to be
\(0.7~\mathrm{mm}\) on average, and standard deviation \(4.4~\mathrm{mm}\). They
also analyzed some factors that can influence GOME-2 retrieval performance. It
was found that increasing solar zenith angle also increases both bias and
standard deviation, that cloud presence increases bias but does not have an
effect on standard deviation, and that, under cloudless conditions, GOME-2 errors
are within its nominal error. Other studies followed in the Iberian Peninsula,
like those by Vaquero-Martinez et al. \cite{Vaquero-Martinez2017c,Vaquero-Martinez2017d,Vaquero-Martinez2018c}, who compared Ozone Monitoring Instrument (OMI), MODIS
and a set of satellite instruments, respectively, against reference GNSS data. We
will focus on the last article, as it includes OMI and MODIS (both NIR and IR products), and also other products like GOME-2,
SCanning Imaging Absorption SpectroMeter for Atmospheric CHartographY
(SCIAMACHY), AIRS, and Spinning Enhanced Visible and Infrared Imager (SEVIRI).
They studied the pseudomedian (rather than mean bias) and the inter-quartile
range (rather than standard deviation) of relative differences between satellite
products and reference GNSS product in nine stations the interior Iberian
Peninsula during the period 2007-2013. It was found that all satellite products
tended to smooth IWV values (low IWV values tended to be overestimated, while high IWV
ones tended to be underestimated). \(R^2\) was in the range \(0.56 ~(\text{AIRS})-0.83~(\text{GOME-2})\), pseudomedians between \(-5.2\pm 0.1\%~(\text{SEVIRI})\) and \(16.7\pm 0.8\%~(\text{GOME-2})\), and inter-quartile
ranges between \(32.58\%~(\text{GOME-2})\) and \(47.84\%~(\text{AIRS})\).
Dependencies on some variables were also studied, namely, solar zenith angle,
IWV, season and cloudiness, which highlighted the differences among the
different satellite remote sensing approaches.

A similar multi-comparison including AIRS, GOME-2, MODIS (NIR and IR), OMI,
POLarization and Directionality of the Earth's Reflectances (POLDER) and
SCIAMCHY against GNSS data in the Ny-Alesund station (European Arctic) during
the period 2010--2017 was performed in another work \citep{Vaquero-Martinez2020c}. It
was found that NIR had a very low performance in this region, and that some
external source for filtering out cloudy scenes is needed. OMI version 4.0
product showed high bias and variability as compared with the other products.
The biases were in the range \(-2.7\)--5.9~\(\mathrm{mm}\), RMSE 2.4--5.2~\(\mathrm{mm}\)
and \(R^2\) between \(0.22\) and \(0.73\).

OMI IWV product has also been validated in successive versions of its
algorithm retrieval \citep{Wang2016d, Wang2019e}. We will only discuss the latest
version (4.0) of the algorithm, which was tested against GNSS IWV
globally distributed network for land and SSMI/S over the oceans. The influence
of cloud scenes is studied, recommending filtering data with cloud fraction less
than 0.05--0.25 and cloud top pressure more than \(750~\mathrm{mm}\), apart from
quality flag. In the comparison, cloud fraction limit is set to \(0.05\), with
bias of \(0.32~\mathrm{mm}\), standard deviation of \(5.2~\mathrm{mm}\) and
correlation \(0.87\) with respect to GNSS. They also demonstrate the usefulness of
OMI data for assimilation with WRF. However, on a more local scale,
\citet{Vaquero-Martinez2020c} showed that, in the Arctic, OMI-GNSS bias can be very high.
Therefore, although OMI product has a good performance globally, it is important
to study local performance when using it for specific regions.

\citet{Ningombam2016a} validated MODIS-NIR IWV against a GNSS station in the
Trans-Himalayan region. Bias was \(-0.18~\mathrm{mm}\) and RMSE
\(1.37~\mathrm{mm}\), while correlation was \(0.95\) for daily data. The
comparison was also performed in a monthly/seasonal basis, finding that summer
values were worse than in other seasons, which is expected because IWV
increases in summer. These values were better than others reported in the
literature for this product, probably because of the use of daily data.

Orbiting Carbon Observatory-2 IWV was compared against the SuomiNet global network
and other instruments (IGRA's radiosondes and
AERONET sun-photometers) as \mbox{references \citep{Nelson2016}}. RMSE value reported
against SuomiNet is \(1.3~\mathrm{mm}\) and correlation of \(0.994\).

\citet{Gong2019} analyzed the Fengyun-3A Medium resolution spectral imager
(FY-3A/MERSI) IWV product against GNSS and sun-photometer measurements in the
Tibetan Plateau in 2009-2013. RMSE was in the range 1.87--2.75~\(\mathrm{mm}\)
and bias $-1.14$--0.64~\(\mathrm{mm}\). \citet{He2019} compared FY-3A/MERSI, Terra/MODIS
and Envisat/MERIS (all NIR satellite instruments) against GNSS in US. It was
found that NIR instruments tend to underestimate IWV under cloudy conditions,
as NIR cannot penetrate clouds. RMSE was of \(8.644~\mathrm{mm}\),
\(5.480~\mathrm{mm}\) and \(3.708 ~\mathrm{mm}\). RMSE was found to increase under
moist conditions. Under clear-sky conditions, \(R^2\) was of \(0.93\) and \(0.95\)
for MERIS and MODIS, and \(0.80\) for MERSI.

\citet{CarbajalHenken2020a} validated the satellite instruments, Infrared Atmospheric
Sounding Interferometer (IASI), MIRS, MODIS, and MODIS-FUB, against the GNSS
network in Germany. The biases were \(1.77\), \(1.36\), \(1.11\), and \(- 0.31~\mathrm{mm}\), respectively, while RMSE were in the range
2.11--3.77~\(\mathrm{mm}\) and correlations  0.88--0.96. Monthly means showed
similar differences, and no impact was found on time-averaging or low temporal
sampling from satellite instruments. Therefore, sampling of IWV seemed to affect
the mean values only marginally, but it could have an effect on frequency
distribution. Nevertheless, in more challenging regions, satellite estimates
of water vapor can improve greatly when averaged in time, as shown by
\citet{Alraddawi2018b}. In this work, monthly IWV products from GOME-2, MODIS and
SCIAMACHY were compared against GNSS in three Arctic stations (in the period
2001--2014). The Arctic is known to be a challenging region for satellite
retrieval, as high solar zenith angle, high frequency of cloudy skies and presence of snow
are problematic conditions for retrieval and remote sensing algorithms.
Generally, seasonal biases were less than \(1~\mathrm{mm}\) for AIRS,
\(2.5~\mathrm{mm}\) for SCIAMACHY, and \(3.5~\mathrm{mm}\) for MODIS. Moreover, the
increase in cloud fraction was related to increasing dry bias. Therefore,
authors suggested that more robust information about clouds must be included in the
retrievals to reduce the aforementioned biases.

The wide variety of satellite products using different instruments and
algorithms, together with different quality criteria and the use of different
regions for testing, makes it very difficult to extract common conclusions to
different works. Even the most used statistical parameters can largely change
with different methodological approaches, especially different quality controls.
For that reason, we conclude that the scientific community should establish a
series of common practices to perform validations in order to make different
works comparable. Moreover, it can be concluded that satellite measurements
can have very good quality if the right conditions are met. There is a general
agreement that cloudy and/or high moisture situations tend to cause worse values
of accuracy and precision metrics (i.e.,~MBE or SD).

\hypertarget{sp}{%
\section{Spatio-Temporal Analysis}\label{sp}}

Table~\ref{tab:spt} summarizes some of the findings of each paper from this section.

\hypertarget{asia}{%
\subsection{Asia}\label{asia}}

\textbf{\hl{China}}. %Is the bold necessary?
 \citet{Jin2008} investigated the spatial, seasonal and diurnal distribution
of water vapor for three years (2004--2007). The region with the highest IWV was the
South-East, with a strong annual cycle (amplitude of $\sim$\(15~\mathrm{mm}\)),
while the lowest IWV was found in the North-West region, with a smaller annual
cycle (amplitude of $\sim$\(4~\mathrm{mm}\)). Mean diurnal cycle amplitude was
found to be \(0.7~\mathrm{mm}\). The spatial distribution was dependent on
latitude, topography, season and monsoon. \citet{Wu2020} studied midsummer diurnal
variations over 18 stations around the Poyang Lake during three year period
(2015--2018). Diurnal IWV typically peaks at 1200--2000 Local Solar Time (LST).
Air temperature diurnal cycle was reported to be likely behind the IWV diurnal
cycle. The semi-diurnal cycle was found to account for about \(50\%\) of variance
contribution. Tibetan Plateau is a region of high interest in China, specially
because of its role in the summer monsoon. \mbox{\citet{Takagi2000} studied} the premonsoon
and monsoon diurnal variability at the station of Lhasa in this region.
Premonsoon minima and maxima occurred at 1800 and 0400 LST, respectively, while,
in the monsoon period, these were displaced to 1500 and 2300 LST. This result
shows that water vapor is affected by local circulation. It also indicates the
usefulness of GNSS for studying the IWV daily cycle and recommends the addition of
more stations to the Plateau in order to study moisture transport in the region.

\textbf{\hl{Japan}}. \citet{Iwasaki2001} studied the diurnal variability in 5 GNSS stations and
the other three radiosonde stations. The maximum was found around 1800--2000 LST, some hours
later than in China. \citet{Li2008} also analyzed the diurnal variations in central Japan
during August 2000, which found the maximum at the same hours. The moisture
transport was found to cause the differences in phase of the diurnal cycle of
IWV among different locations.

\textbf{\hl{Sumatra}}. \citet{Wu2003} presented the diurnal variation in the island of Sumatra,
and found a distinct diurnal cycle. \citet{Torri2019} used data from the Sumatran GNSS
array to study the diurnal cycle of water vapor. The Madden--Julian oscillation
(MJO) was found to impact on the diurnal cycle, appearing smaller in both the daily
mean and the amplitude of the diurnal cycle during the suppressed phase relative
to the developing/active MJO phase. The evening/nighttime peaks of IWV
offshore appeared later during the suppressed phase of the MJO compared to its
active phase (from 1900 LST in the active phase to \mbox{2200 LST} in the suppressed
phase).

\textls[25]{\textbf{\hl{South Korea}}. \citet{Sohn2010} estimated the trend over 5 stations of South Korea
\mbox{(2000--2009)} as \(0.11~\mathrm{mm}\)~{year}$^{-1}$. They also calculated seasonal
trends for summer (\(0.16~\mathrm{mm}\)~{year}$^{-1}$), winter
(\(0.04~\mathrm{mm}\)~{year}$^{-1}$), spring (\(0.05~\mathrm{mm}\)~{year}$^{-1}$), and
autumn ($-0.14~\mathrm{mm}$~{year}$^{-1}$).}
\newpage

\end{paracol}
\nointerlineskip
\begin{specialtable}[H]
\widetable
\caption{\label{tab:spt}Summary of spatio-temporal studies. Unless indicated otherwise, times are given in local solar time and trends in $\mathrm{mm~{year}^{-1}}$.}
\scalebox{0.92}[0.92]{

\begin{tabular}{m{0.8cm}<{\centering}m{1cm}<{\centering}m{0.6cm}<{\centering}m{1.5cm}<{\centering}m{1.2cm}<{\centering}m{3.5cm}<{\centering}m{2cm}<{\centering}m{5.8cm}<{\centering}}
\toprule
\textbf{Work} & \textbf{Year} & \textbf{N} & \textbf{Start} & \textbf{End} & \textbf{Region} & \textbf{Type} & \textbf{Value}\\
\midrule
\citep{Iwasaki2001} & 2001 & 5 & \hl{1999-07} %MDPI: please check if this is year and month.
 & 1999-08 & Japan & Diurnal & max: 1900\\
\midrule
\citep{Dai2002} &  & 54 & 1996-01 &  &  {North America (US)} & Diurnal & {max: 1000--1400  (winter) and midafternoon-midnight (summer)}\\
\cmidrule{1-1}
\cmidrule{3-4}
\cmidrule{6-8}
\citep{Gradinarsky2002} & \multirow[t]{-2}{*}{\raggedleft\arraybackslash 2002} & 17 & 1993-08 & \multirow[t]{-2}{*}{\raggedright\arraybackslash 2000-12} &  {Scandinavia (Sweden)} & Trends & 0.15\\
\midrule
\citep{Combrink2007} & 2007 & 2 & 1998-01 & 2006-12 &  {Africa (South Africa)} & Trends & non-significant\\
\midrule
\citep{Li2008} & 2008 & 6 & 2000-08 & 2000-08 & Japan & Diurnal &  {max: 1900, min: noon}\\
\midrule
\citep{Jakobson2009} &  & 32 & 1996-01 &  &  {Baltic (Finland, Latvia, Sweden)} & Diurnal & -\\
\cmidrule{1-1}
\cmidrule{3-4}
\cmidrule{6-8}
 &  &  &  &  &  & Seasonal & \\
\cmidrule{7-8}
\multirow[t]{-2}{*}{\raggedright\arraybackslash \citep{Suparta2009a}} & \multirow[t]{-3}{*}{\raggedleft\arraybackslash 2009} & \multirow[t]{-2}{*}{\raggedleft\arraybackslash 9} & \multirow[t]{-2}{*}{\raggedright\arraybackslash 2003-01} & \multirow[t]{-3}{*}{\raggedright\arraybackslash 2005-12} & \multirow[t]{-2}{*}{\raggedright\arraybackslash Antarctic} & Diurnal &  {max: 0600--0900 UTC, min: 1800--2100 UT}\\
\midrule
\citep{Sohn2010} & 2010 & 5 & 2000-01 & 2009-12 & SouthKorea & Trends & 0.11\\
\midrule
\citep{OrtizdeGalisteo2011c} & 2011 & 10 & 2002-01 & 2011-12 &  {Iberian P. (Spain)} & Diurnal & {min: 0430--0530 UTC, Afternoon}\\
\midrule
 &  &  &  &  &  & Seasonal & \\
\cmidrule{7-8}
\multirow[t]{-2}{*}{\raggedright\arraybackslash \citep{Sharifi2015}} & \multirow[t]{-2}{*}{\raggedleft\arraybackslash 2015} & \multirow[t]{-2}{*}{\raggedleft\arraybackslash 4} & \multirow[t]{-2}{*}{\raggedright\arraybackslash 2009-01} & \multirow[t]{-2}{*}{\raggedright\arraybackslash 2012-12} & \multirow[t]{-2}{*}{\raggedright\arraybackslash Europe} & Diurnal & -\\
\midrule
\citep{Alshawaf2017a} &  & 113 & 1996-01 & 2015-12 &  {Europe (Germany)} & Trends & $-$0.15--0.23\\
\cmidrule{1-1}
\cmidrule{3-8}
 &  &  &  &  &{Northeast India (India)} & Seasonal &  {max: summer,  min: winter}\\
\cmidrule{6-8}
\multirow[t]{-2}{*}{\raggedright\arraybackslash \citep{Barman2017}} & \multirow[t]{-3}{*}{\raggedleft\arraybackslash 2017} & \multirow[t]{-2}{*}{\raggedleft\arraybackslash 5} & \multirow[t]{-2}{*}{\raggedright\arraybackslash 2004-01} & \multirow[t]{-2}{*}{\raggedright\arraybackslash 2012-12} &  {Northeast India (India)} & Diurnal & max: 1000--1300 UTC\\
\midrule
 &  &  &  &  & {Central France (France)} & Diurnal & -\\
\cmidrule{6-8}
\multirow[t]{-2}{*}{\raggedright\arraybackslash \citep{Hadad2018}} & \multirow[t]{-2}{*}{\raggedleft\arraybackslash 2018} & \multirow[t]{-2}{*}{\raggedleft\arraybackslash 2} & \multirow[t]{-2}{*}{\raggedright\arraybackslash 2006-01} & \multirow[t]{-2}{*}{\raggedright\arraybackslash 2018-01} &  {Central France (France)} & Trends & non-significant\\
\midrule
\citep{Torri2019} &  & 24 & 2005-01 & 2008-12 & Sumatra & Diurnal &  {max: 1900 (land); 1300 and 0100 (offshore)}\\
\cmidrule{1-1}
\cmidrule{3-8}
 &  &  &  &  &  & Seasonal & \\
\cmidrule{7-8}
\multirow[t]{-2}{*}{\raggedright\arraybackslash \citep{Zhao2019b}} & \multirow[t]{-3}{*}{\raggedleft\arraybackslash 2019} & \multirow[t]{-2}{*}{\raggedleft\arraybackslash 192} & \multirow[t]{-2}{*}{\raggedright\arraybackslash 2004-01} & \multirow[t]{-2}{*}{\raggedright\arraybackslash 2017-12} & \multirow[t]{-2}{*}{\raggedright\arraybackslash Global} & Trends & -\\
\midrule
\citep{Bernet2020} &  & 35 & 2000-01 & 2018-12 &  {Europe (Switzerland)} & Trends & 0.001--0.109\\
\cmidrule{1-1}
\cmidrule{3-8}
\citep{Bousquet2020} &  & 8 & 2018-06 & 2020-03 & SW Indian Ocean & Diurnal & -\\
\cmidrule{1-1}
\cmidrule{3-8}
 &  &  &  &  &  & Seasonal & max: austral summer\\
\cmidrule{7-8}
\multirow[t]{-2}{*}{\raggedright\arraybackslash \citep{Lees2020a}} &  & \multirow[t]{-2}{*}{\raggedleft\arraybackslash 12} & \multirow[t]{-2}{*}{\raggedright\arraybackslash 2009-01} & \multirow[t]{-2}{*}{\raggedright\arraybackslash 2019-12} & \multirow[t]{-2}{*}{\raggedright\arraybackslash Indian Ocean} & Diurnal & {max: late afternoon/evening (land); night (ocean); early morning (coastal), min: morning (land); mid-day (ocean); late afternoon (coastal)}\\
\cmidrule{1-1}
\cmidrule{3-8}
 &  &  &  &  &  & Seasonal & max: mid-May--end-October\\
\cmidrule{7-8}
\multirow[t]{-2}{*}{\raggedright\arraybackslash \citep{Trakolkul2020a}} &  & \multirow[t]{-2}{*}{\raggedleft\arraybackslash 11} & \multirow[t]{-2}{*}{\raggedright\arraybackslash 2007-01} & \multirow[t]{-2}{*}{\raggedright\arraybackslash 2015-12} & \multirow[t]{-2}{*}{\raggedright\arraybackslash Thailand} & Trends & -\\
\cmidrule{1-1}
\cmidrule{3-8}
\citep{Wu2020} & \multirow[t]{-7}{*}{\raggedleft\arraybackslash 2020} & 18 & 2015-01 & 2018-12 & {Poyang Lake (China)} & Diurnal & max: 1600\\
\bottomrule
\end{tabular}}
\end{specialtable}
\begin{paracol}{2}
%\linenumbers
\switchcolumn

\textbf{\hl{India}}. \citet{Barman2017} analyzed GNSS IWV from five stations in India during the
period 2004--2012. They found that high altitude sites had weaker annual
variability than low altitude ones. However, seasonal variability was similar in
the five sites. Seasonal IWV was strongly linked to the monsoon, with maximum in summer
and minimum in winter. The diurnal cycle was marked in high altitude sites, while
weak in low altitude ones. \mbox{\citet{Bousquet2020} presented} a new GNSS network in the
Southwest Indian Ocean in the frame of the Indian Ocean GNSS Application for
Meteorology Project, which is being deployed since November 2017. They analyzed
the period from 2018 to 2020 and found that minimum seasonal values decrease away
from the equator. Diurnal cycle was found to differ at both sides of Madagascar.
This study was extended by \citet{Lees2020a}, where the seasonal time-scale was
studied. The annual amplitude of IWV was reported to vary in the interval 10--15 ~\(\mathrm{mm}\) near the equator and in the interval 20--30 ~\(\mathrm{mm}\) in the
subtropics. The strongest daily IWV amplitudes were found in the mountainous
islands. The reanalysis ERA5 was found to reproduce the main features of IWV
time series.

\textbf{\hl{Thailand}}. \citet{Trakolkul2020a} studied GNSS IWV series of 11 CORS stations in
Thailand to analyze seasonal features and trends. The annual variation amplitude
was reported to be in the range 6--19 ~\(\mathrm{mm}\).

\hypertarget{europe}{%
\subsection{Europe}\label{europe}}

\textbf{\hl{General}}. \citet{Sharifi2015} used a hybrid approach to time series modeling of GNSS
IWV series over four sites in Europe (years 2009--2012, 5-min sampling), combining
least squares harmonic estimation (for modeling harmonic characteristics) with
least squares support vector machine optimized by cross-validation strategy (for
modeling non-harmonic features). Inter-seasonal variability was found to be greater
than inter-annual variability. Diurnal cycles were strong in summer with maxima
between 1200 and \mbox{1800 LST}. The peaks were related to solar flux and the phase to
other features, such as location or \mbox{wind/temperature patterns.}

\textbf{\hl{Northern Europe}}. \citet{Gradinarsky2002} studied trends of water vapor in the
region of Scandinavia during the years 1993 and 2000. They found that the GNSS
station radome can affect measurements to an extent between \(0.0\) and \(1.8~\mathrm{mm}\).
Trends were positive, \mbox{0.1--0.2~\(\mathrm{mm~year^{-1}}\)}. These GNSS-calculated
trends were found to be consistent with other measurements (i.e.,~radiosondes and
MWR), assessing the potential of GNSS IWV in climate monitoring. \citet{Jakobson2009}
analyzed the diurnal variability of water vapor in \mbox{32 sites} in the Baltic
region, during 1996--2005. Peak to peak difference was of 0.5--0.6 ~\(\mathrm{mm}\)
in spring/summer, while winter and autumn lacked a clear pattern. The high
variability (mainly with synoptic situation, or substitution of air masses)
makes that the diurnal cycle can only be observed when averaging over many years.

\textbf{\hl{France}}. \citet{Hadad2018} combined different types of water vapor measurements to
study two sites in France. The work presented a non-significant trend at the
site of Cézeaux of \(0.42 \pm 0.45 ~\mathrm{mm~decade^{-1}}\). Both stations
presented almost no diurnal cycle and the authors concluded that variability of
surface water vapor was influenced by sporadic circulation patterns rather
than by regular diurnal variations.

\textbf{\hl{Switzerland}}. \citet{Bernet2020} studied trends of water vapor over Switzerland
using different instruments, including 31 GNSS sites in the period 1995 to 2018.
Jumps caused by instrumentation changes were corrected. IWV was found to
increase with a rate of \(2-5 \%~\mathrm{decade^{-1}}\). From the comparison among different
instruments, it was concluded that GNSS data are reliable for the detection of
trends, but a few stations need quality control and harmonization.

\textbf{\hl{Spain}}. \citet{OrtizdeGalisteo2011c} analyzed the diurnal cycle of IWV in 10 GNSS
Spanish stations for the years 2002-2011, finding that it explained 3\%--7\%
of sub-daily annual variability. IWV was found to peak at 0430--0530 UTC, and the
diurnal cycle was found to be more important in the Mediterranean sites than in
the inland or Atlantic ones. The summer diurnal cycle is also more variable among sites.
The first two harmonics (diurnal and semidiurnal cycles) explained \(97\%\) of
the variance. Another work analyzed the annual cycle in the same period and
stations
\citep{OrtizdeGalisteo2014b}, finding that the 12-month harmonic explained \(96\%\) of
variance, with a maximum in summer and minimum in the winter months. Southwestern
sites, however, showed a local minimum in July, which can be explained by the
presence of dry air masses. The Mediterranean sites had larger amplitudes than
inland sites. The July local minimum had been already reported by \citet{Torres2010},
who characterized the IWV in the Gulf of Cadiz region, in southern Iberian
Peninsula using GNSS, radiosondes and a sun-photometer.

\textbf{\hl{Germany}}. \citet{Alshawaf2017a} studied the IWV time series over 113 GNSS stations in
Germany, with 10--19 years of temporal coverage, together with meteorological
data from automatic stations and ERA-Interim reanalysis. GNSS IWV trend ranged
between \(-1.5\) and \(2.3~\mathrm{mm~decade^{-1}}\). However, longer time-series
from ERA-Interim (\(\sim\)\(26~\mathrm{year}\)) revealed positive trends in all
sites, which also had a gradient from southwest to northeast. Trends obtained
from dew-point temperature were comparable with the other estimates, and
therefore can be used when observations are not available.

\hypertarget{africa}{%
\subsection{Africa}\label{africa}}

\citet{Combrink2007} studied times-series of GNSS-derived IWV in South Africa, with the aim of detecting trends in the period. They used the autoregressive--moving-average (ARMA)(1,1) model, concluding that this model represents better the real trend, provided that more than five years of data are used. They also concluded that none of the two stations used had statistically significant trends.

\hypertarget{america}{%
\subsection{America}\label{america}}

\citet{Dai2002} studied GNSS IWV from 54 stations over North America. Diurnal variations were significant, explaining over the \(50\%\) of the subdaily variance, specially in summer, with weaker diurnal cycles in the rest of the seasons. Peaks are found around 1000--1400 LST in winter, and from midafternoon to to midnight in summer. The semidiurnal cycle is rather weak.

\citet{Yu2017a} proposed a model to spatially interpolate GNSS ZTD and used interpolated values from meteorological stations for pressure and \(T_m\) in California. The interpolation method is iterative and tries to decouple turbulent- and elevation-dependent ZTD components. Validation with MODIS as a reference yielded RMSE of \(1.70~\mathrm{mm}\).

\citet{Bordi2014} studied temporal series of six GNSS stations in Eastern US, where it was found that the variability of IWV is dominated by the annual cycle, with a missing daily cycle, while RH from weather stations is more characterized by the daily cycle, without an annual cycle.

\hypertarget{antarctic-region}{%
\subsection{Antarctic Region}\label{antarctic-region}}

\citet{Suparta2009a} studied the GNSS IWV over 9 sites in Antarctic during two years (2003--2005). The diurnal variation was remarkable in summer, but not in winter. IWV was higher in summer than in winter. Moreover, it was found that the presence of strong winds can cause a decrease in monthly IWV.

\hypertarget{global}{%
\subsection{Global}\label{global}}

\citet{ShuanggenJin2009} analyzed 13 years (1995--2007) of data from 155 GNSS stations globally and distributed and studied the seasonal variation of IWV. Globally, seasonal maxima were found in summer and minima in winter. Annual cycle amplitude was \(10 – 20 \pm 0.5~\mathrm{mm}\) in mid-latitudes and \(5 \pm 0.5 ~\mathrm{mm}\) in high latitudes and equatorial areas. Semiannual variation amplitudes were less important, at about \(0.5 \pm 0.2 ~\mathrm{mm}\). The diurnal cycle was also studied and it was found to have an amplitude of \(0.2 – 1.2 \pm 0.1 ~\mathrm{mm}\), and the maximum and minimum occurred around noon and midnight, respectively. The semidiurnal cycle was reported to have a weaker amplitude below \(0.3~\mathrm{mm}\).
\mbox{\citet{Zhao2019b} agreed} with this, finding that annual variability was the more important component in IWV variation. This last work used 192 IGS stations distributed globally to interpolate and analyze IWV at a global scale during the period 2004--2017. Despite the different periods with very few years in common, both works show similar results.

Regarding trends, \citet{Chen2016b} presented a global analysis of the trends of IWV using different products, including reanalysis ERA-Interim and NCEP, SSM/I microwave satellite, radiosondes and GNSS. Particularly, 100 GNSS stations were used in the study, with a time-span of 15 years. The other products had longer time-series, reaching \mbox{36 years} of reanalysis and radiosondes. Therefore, the trends were calculated at different timescales to allow comparisons. The 15 year period 2000--2014 was concluded to have stronger upward trends than longer periods. The trends were not homogeneous, being negative in Asia, the United States or North Africa, while large positive trends were found in the North and West Pacific, Europe, the eastern Indian Ocean, and southern Asia. The polar region is the one with a larger increase in IWV. Another interesting issue discussed in this work is the relationship between surface temperature and IWV. According to Clausius-Clapeyron equation, the increase rate of water vapor should be 6\%--13\%\(\mathrm{K^{-1}}\). However, some regions depart greatly from this range, reaching negative values in this rate. Such large departures were associated to regional moisture divergence/convergence. At a global scale, the Clausius-Clapeyron estimate remains valid. This issue was also studied by \mbox{\citet{Wang2016e}, who} found that using night surface temperature had better correlation with IWV, and therefore it was suggested to use it for studies on water vapor feedback.

\hypertarget{impact}{%
\section{Impact of Water Vapor Variability on Meteorology and Climate}\label{impact}}

\hypertarget{assimilation}{%
\subsection{Assimilation}\label{assimilation}}

NWP models need to ingest real data in the process known as assimilation to be able to perform predictions. Therefore, GNSS data can be an interesting source for NWP assimilation, especially in regions where other sources of atmospheric water vapor information are missing. \citet{Gutman2001} showed that GNSS data improved forecasts from NWP models.
Particularly they used the Rapid Update Cycle in the US, and assimilated data from 18 and 55 stations.
They found improvements in relative humidity (RH) of up to \(4.5\%\) for the 55-station experiment and up to \(1.0\%\) for the 18-station experiment.
Despite being a modest improvement, it must be taken into account that the US had already good knowledge of the moisture from other kinds of measurements already assimilated by the model.
Therefore, in other regions this could be enhanced, and even in the US, the improvement was evident, especially in the 3-h forecast.
They also developed techniques to retrieve GNSS IWV measurements every 30 min to be available for forecasters and modelers, improving IGS service in latency. Other works from the 2000s found moderated improvements in models as well: \citet{Guerova2004}, who assimilated GNSS IWV from COST Action 716 into the MeteoSwiss NWP model; \citet{Yuan2005}, who assimilated IWV data from 11 GNSS stations in China into the MM5 model; \citet{Gendt2004}, who presented a near real-time GNSS IWV product in the context of the project GASP and assimilated it into the local model from German Weather Service; and \citet{Zhang2006}, who tested the GNSS assimilation in a severe weather event in Beijing.

All these works considered the assimilation of the IWV GNSS data.
However, it is possible to use more direct data, such as SWD.
\citet{Ha2003} compared the assimilation of SWD and IWV from the GNSS network and found SWD to be more appropriate for correctly retrieving temperature and moisture profiles.
They used a hypothetical GNSS network of 64 stations in the US, using the MM5 4-dimensional variational data assimilation system.
\citet{Liu2006} used a similar approach, but assimilated slant water vapor (SWV) into the 3D-variational (3DVAR). The tests included hypothetical GNSS stations as well and sensitivity analysis, concluding that GNSS SWV assimilation is positive even with the inclusion of typical errors in measurements.

The introduction of multi-GNSS processing can improve near real-time measurements in several ways.
First, it can provide more accurate data by increasing the number of frequencies used, as new GNSS satellites introduce a third frequency (better combinations of signals). Second, having more satellites available makes possible to increase temporal resolution (below 30 min) and spatial resolution through networks and tomography techniques. \citet{Karabatic2011} performed an experiment in Austria with the use of multi-GNSS (GNSS and GLONASS) network, finding the assimilation of multi-GNSS data very positive and concluding that Galileo constellation inclusion could improve analysis even more.

In addition to this, it must be noted that regions lacking dense networks of ground-based GNSS stations can also benefit from GNSS data assimilation. \citet{Kumar2017} showed that assimilating GNSS data from a single station was enough to locally improve ($\sim$\(10\%\))  the short-term forecasts in an experiment over southern India during the monsoon. Of course, it will be better if there is moisture information from more GNSS stations, but regions lacking moisture information from other sources can benefit from GNSS data even with a single station in the region of interest.
If two-frequency stations are not available, it may be possible to use single-frequency receivers for assimilation. This issue is still under study and more research could be useful, as the use of single-frequency receivers can provide very dense networks as compared to dual-frequency ones. One experiment was performed using the Regional Atmospheric Modeling System (RAMS) 3D variational data assimilation system to assimilate GNSS-ZTD from single-frequency receivers. It was concluded that, although results are preliminary and some issues must be addressed, the assimilation was very positive for both ZTD and IWV fields \citep{Mascitelli2019}.

\hypertarget{circulation}{%
\subsection{Circulation}\label{circulation}}

As pointed out in Section~\ref{ic}, water vapor is greatly affected by circulation patterns, which can transport moisture or dry air to different regions. For this reason, GNSS can provide information about these circulation patterns in certain regions.
\citet{Graham2012} used GNSS IWV data to show that important differences in IWV can
occur between the Swiss plains and nearby Alps during some thunderstorms events,
attributable to regional airflow convergence. They presented a method to
acquire information about the amount and change of total IWV over time for the
mesoscale in the Bernese Oberland area using the GNSS IWV data. An
anti-correlation pattern was found between Payerne and Saanen, which suggested a
mountain-plain circulation pattern. Moreover, \citet{Kingsmill2013} and \citet{Neiman2013a} used GNSS IWV products among other remote-sensing
information about the state of the atmosphere to study the northern California
Valley, which has sierra barrier jets and land-falling atmospheric rivers
(ARs), modulators of the rainfall in this region. The study links
these two phenomena with the orographic configuration of the Valley, and the
concurrence of both phenomena at the same time, or just one of them, affects the spatio-temporal rainfall pattern.

GNSS has been used to study water vapor during El Niño events in the coast of the South China Sea \citep{Suparta2018}. During these events, IWV drops, reaching a minimum about 10 weeks later, while the opposite happens during La Niña events. Therefore, GNSS IWV is suggested to be a useful instrument for studying ENSO activity.

\hypertarget{radiative-transfer-studies}{%
\subsection{Radiative Transfer Studies}\label{radiative-transfer-studies}}

There has been recent interest in the role of water vapor in radiative
balance, enhanced by the increase in water vapor data availability, with higher
spatial and temporal resolution. \citet{Vaquero-Martinez2018a} studied the short-wave
effects of water vapor using 20 GNSS stations in the Iberian Peninsula as input
to Santa Barbara's dissort % please check this is the correct word
radiative transfer model (SBDART). They found that
the effect for cloudless skies followed an empirical power relation of the
short-wave effect with IWV and the cosine of the solar zenith angle. An extended study
\citep{Vaquero-Martinez2020}, including short- and long-wave in seven selected
stations, found that positive trends for long-wave (and total) water vapor
effects could partially explain the well-known increase of surface air
temperatures in the Iberian Peninsula, while the negative trend in the
short-wave effect represented about a quarter of the contribution of aerosols to
the so-called brightening effect.

In a more practical fashion, GNSS data are also used for the modeling of atmospheric
transmission. The PWV\_KPNO python package \citep{Perrefort2019} uses GNSS IWV data
from the SuomiNet network for modeling atmospheric transmission function due to
water vapor. The package interpolates IWV data from the GNSS stations to the desired
position and then applies the MODTRAN model to obtain the atmospheric transmission.

\hypertarget{conclusions}{%
\section{Conclusions}\label{conclusions}}

GNSS sensing of atmospheric water vapor is already a well established technique that allows for the monitoring of this trace gas with wide (yet land-only) coverage, with high temporal resolution and without weather restrictions. These features make this technique very suitable for testing other techniques, especially when weather is adverse. In comparison with other products, GNSS water vapor data are high quality, being considered a reference for the validation of other ground-based and satellite-based water vapor measurements and model estimates. However, we consider that special care must be taken with some issues. First, IWV is dependent on height and therefore when comparing data related to different heights (i.e., ground-based stations and different heights, satellite or NWP pixels at a mean height of the area, different from the real height of the GNSS station) it is important to apply the corresponding correction to account for this height difference. Second, the homogeneity of the series must be checked (especially if there are changes in antenna or processing strategy). Third, it is considered a rule of thumb that GNSS retrieve water vapor information of about \(100~\mathrm{km}\) radius, which can be used to set a condition on distance to GNSS station when performing validations.

Nowadays, time-series of GNSS data products are starting to be long enough to study water vapor in long term studies, such as its trends and variability across different time scales. Additionally, the increasing amplitude of the ground-based GNSS networks allows for a detailed analysis of its spatial distribution. It is observed that, despite the average behavior based mainly on temperature (according to the Clausius-Clapeyron equation), local circulation is more important. Although there are clear cycles in daily and annual regimes, situations with relevant local circulation can disrupt these. Therefore, it is fundamental to use long-term data to study daily and annual cycles, especially in regions where strong circulation patterns can modify air moisture.

GNSS water vapor has been widely used for several meteorological and climate applications such as assimilation, circulation of the atmosphere and radiative effects. It is shown that assimilation can be improved with GNSS data, especially in regions where there may be a lack of moisture information. It must be noted that GNSS are relatively cheap and easy to maintain, and provide continuous measurements for all-weather conditions. This means they can be used in regions with fewer resources or those where human access \mbox{is difficult.}

Despite being a well established technique, it must be noted that many other GNSS-based techniques, such as tomography or data fusion approaches, can make the best out of the information that GNSS meteorology has to offer. The use of multi-GNSS techniques, especially, have the potential to improve the quality of these approaches. The use of low-cost devices (i.e., single frequency receivers) can be interesting in terms of retrieving moisture information cost-effectively.

In summary, the GNSS technique allows the retrieval of accurate atmospheric water vapor data. This method exhibits great potential to be extraordinarily useful in the monitoring of the weather and climate.
%%%%%%%%%%%%%%%%%%%%%%%%%%%%%%%%%%%%%%%%%%

\vspace{6pt}

%%%%%%%%%%%%%%%%%%%%%%%%%%%%%%%%%%%%%%%%%%
%% optional
%\supplementary{The following are available online at \linksupplementary{s1}, Figure S1: title, Table S1: title, Video S1: title.}


%%%%%%%%%%%%%%%%%%%%%%%%%%%%%%%%%%%%%%%%%%
\authorcontributions{Conceptualization, M.A.; Data curation, J.V.-M.; writing---original draft preparation, J.V.-M.; writing---review and editing, M.A.; supervision, M.A.~All authors have read and agreed to the published version of the manuscript.}

\funding{This work was partly supported by the Ministerio de Economia y Competitividad of the Spanish Government (CGL2017-87917-P), and by Junta de Extremadura and European Regional Development Fund (IB18092). Javier Vaquero-Martinez was supported by a predoctoral fellowship (PD18029) from Junta de Extremadura and European Social Fund.}

\institutionalreview{\hl{note} %MDPI: please add this part.
}%In this section, please add the Institutional Review Board Statement and approval number for studies involving humans or animals. Please note that the Editorial Office might ask you for further information. Please add ``The study was conducted according to the guidelines of the Declaration of Helsinki, and approved by the Institutional Review Board (or Ethics Committee) of NAME OF INSTITUTE (protocol code XXX and date of approval).'' OR ``Ethical review and approval were waived for this study, due to REASON (please provide a detailed justification).'' OR ``Not applicable'' for studies not involving humans or animals. You might also choose to exclude this statement if the study did not involve humans or animals.}

\informedconsent{\hl{note} %MDPI: please add this part.
}%Any research article describing a study involving humans should contain this statement. Please add ``Informed consent was obtained from all subjects involved in the study.'' OR ``Patient consent was waived due to REASON (please provide a detailed justification).'' OR ``Not applicable'' for studies not involving humans. You might also choose to exclude this statement if the study did not involve humans.

%Written informed consent for publication must be obtained from participating patients who can be identified (including by the patients themselves). Please state ``Written informed consent has been obtained from the patient(s) to publish this paper'' if applicable.}

\dataavailability{\hl{note} %MDPI: please add this part.
}%In this section, please provide details regarding where data supporting reported results can be found, including links to publicly archived datasets analyzed or generated during the study. Please refer to suggested Data Availability Statements in section ``MDPI Research Data Policies'' at \url{https://www.mdpi.com/ethics}. You might choose to exclude this statement if the study did not report any data.} 

\conflictsofinterest{The authors declare no conflict of interest.}

%% Optional
%\sampleavailability{Samples of the compounds ... are available from the authors.}

%%%%%%%%%%%%%%%%%%%%%%%%%%%%%%%%%%%%%%%%%%
%% Only for journal Encyclopedia
%\entrylink{The Link to this entry published on the encyclopedia platform.}

%%%%%%%%%%%%%%%%%%%%%%%%%%%%%%%%%%%%%%%%%%
% Optional
\abbreviations{Abbreviations}{The following abbreviations are used in this manuscript:\\
% I don't know how to insert this so I do it by hand:
\noindent
\begin{tabular}{@{}ll}
AIRS & Atmospheric Infrared sounder\\
AMSU-B & Advanced Microwave Sounding Unit\\
APC & Antenna Phase Center\\
AR & atmospheric river\\
BALTEX & Baltic Sea Experiment\\
BDS & BeiDou Navigation Satellite System (Chinese GNSS)\\
CORS & Continuously Operating Reference Stations\\
COST & Cooperation in Science and Technology\\
ECMWF & European Centre for Medium-Range Weather Forecasts\\
FTIR & Fourier-transform infrared spectroscopy\\
FY-3A/MERSI & Fengyun-3A Medium resolution spectral imager\\
GASP & Atmosphere Sounding Project\\
GLONASS & Globalnaya navigatsionnaya sputnikovaya sistema (Russian GNSS)\\
GNSS & Global Navigation Satellite Systems\\
GOME-2 & Global Ozone Monitoring Experiment 2\\
GPS & Global Positioning System\\
GPT & Global Pressure and Temperature\\
HIRLAM & High Resolution Limited Area Model\\
IGS & International GNSS Service\\
InSAR & Interferometric synthetic-aperture radar\\
IPWV & Integrated precipitable water vapor\\
ITU-R & International Telecommunication Union - Radiocommunications\\
IWV & Integrated water vapor\\
JRA-55 & Japanese 55-year Reanalysis\\
MERIS & MEdium Resolution Imaging Spectrometer\\
MERRA-2 & Modern-Era Retrospective analysis for Research and Applications Version 2\\
MFRSR & Multifilter rotating shadowband radiometer\\
MGEX & Multi-GNSS Experiment\\
MJO & Madden–Julian oscillation\\
MM5 & Fifth-Generation Penn State/NCAR Mesoscale Model\\
NWP & Numerical weather models\\
OMI & Ozone Monitoring Instrument\\
POLDER & POLarization and Directionality of the Earth's Reflectances\\
PW & Precipitable water\\
PWV & Precipitable water vapor\\
RAMS & Regional Atmospheric Modeling System\\
RH & Relative humidity\\
RMSE & Root mean square error\\
SBDART & Santa Barbara's dissort radiative transfer model\\
SCIAMACHY & SCanning Imaging Absorption SpectroMeter for Atmospheric CHartographY\\
SD & Standard deviation\\
SEVIRI & Spinning Enhanced Visible and Infrared Imager\\
SSM/I & Special Sensor Microwave/Imager\\
SSMI/S & Special Sensor Microwave Imager / Sounder\\
STD & Slant tropospheric delay\\
SWD & Slant wet delay\\
T\textsubscript{m} & Weighted mean temperature of the atmosphere\\
T\textsubscript{s} & Surface temperature\\
VMF1 & Vienna Mapping Functions 1\\
WVR & Water vapor radiometer\\
ZHD & Zenith hydrostatic delay\\
ZTD & Zenith tropospheric delay\\
ZWD & Zenith wet delay%\\
\end{tabular}
}
% %%%%%%%%%%%%%%%%%%%%%%%%%%%%%%%%%%%%%%%%%%
% %% Optional
% \appendixtitles{no} % Leave argument "no" if all appendix headings stay EMPTY (then no dot is printed after "Appendix A"). If the appendix sections contain a heading then change the argument to "yes".
% \appendixstart
% \appendix
% \section{}
% \subsection{}
% The appendix is an optional section that can contain details and data supplemental to the main text---for example, explanations of experimental details that would disrupt the flow of the main text but nonetheless remain crucial to understanding and reproducing the research shown; figures of replicates for experiments of which representative data are shown in the main text can be added here if brief, or as Supplementary Data. Mathematical proofs of results not central to the paper can be added as an appendix.
%
% \section{}
% All appendix sections must be cited in the main text. In the appendices, Figures, Tables, etc. should be labeled, starting with ``A''---e.g., Figure A1, Figure A2, etc.
%
% %%%%%%%%%%%%%%%%%%%%%%%%%%%%%%%%%%%%%%%%%%
\end{paracol}
\reftitle{References}

\end{document}